%% file: rpo3.tex
\documentclass[12pt]{article}
\include{head}

\begin{document}
\title{The nature of the continuum limit in the 2D $RP^2$ gauge model}
\author{S.M. Catterall, M. Hasenbusch, R.R. Horgan, R. Renken}
\maketitle
\begin{abstract}
The $RP^2$ gauge model which allows interpolation between the $RP^2$
and $O(3)$ spin models is studied in 2D. We use Monte-Carlo renormalization 
techniques for blocking the mean spin-spin interaction, $\AE$, and the mean
gauge field plaquette, $\PE$. The presence of the $O(3)$ renormalized trajectory
is verified and is consistent with the known three-loop $\gb$-function. The first-order 
`vorticity' transition observed by Solomon et al. \cite{SSD} is confirmed, 
and the location of the terminating critical point is established. New
scaling flows in $(\AE,\PE)$ are observed associated with a large exponent $\k$
in the range $4 \sim 5$. The scaling flows are found to give rise to a strong cross-over 
effect between regions of high and low vorticity and are likely
to induce an apparent signal for scaling in the cross-over region which we propose
explains the scaling observed for $RP^2$ and $RP^3$ models \cite{caea} and also
in a study of the $SO(4)$ matrix model \cite{haho}.  We show that the signal for
this `pseudo' scaling will occur for the $RP^2$  spin model in the cross-over region
which is precisely the region in which computer simulations are done.  We find that
the $RP^2$ spin model is in the same universality class as the $O(3)$ spin model
but that it is likely to require a very large correlation length before the true
scaling of this class sets in. We conjecture that the scaling flows are due either 
to the influence of a nearby new renormalized trajectory or to the ghost of the 
Kosterlitz-Thouless trajectory in the associated $XY$ model. In the former case 
it is argued that the `vorticity' fixed point controlling the critical behaviour 
terminating the first-order line cannot be identified with the conjectured new 
renormalized trajectory.
\end{abstract}
\vfill
\hfill DAMTP 97-68 and HUB-EP-97/78
\newpage
\newsection{\label{intro}Introduction}
The nature of the phase diagram for two-dimensional $RP^N$ models has
been the subject of much recent discussion \cite{caea}, \cite{Sokal}. In
\cite{caea}, Caracciolo et al. compare the correlation length computed
from simulation with that predicted from the perturbative $\gb-$function
using the exact results for the mass gap in $O(N)$ models. They found that
for $RP^2$ ($RP^3$) the observed correlation length on lattices up to $L=512$ was 
smaller than the expected value by a factor of $10^7$ ($10^4$). Their conclusion was 
that either the asymptotic regime is very far indeed removed from the regime of their
study, requiring lattices of sizes of $10^9$ ($10^5$), or that these theories 
were not asymptotically free but that there exists a phase transition at 
finite $\gb$ (non-zero temperature). Caracciolo et al. indeed provide evidence
for the latter scenario by showing that their data scale in a manner consistent with
a Kosterlitz-Thouless parametrization. The two persuasive features are thus that
the correlation length is much smaller than that expected assuming an asymptotically 
free theory and that scaling of the data is observed. This phenomenon occurs in a 
large class of models and the question is whether the signal for a phase transition at
finite $\gb$ and the observed scaling of data are genuine or not. 

The same effects have been observed to a less extreme extent by Hasenbusch and Horgan 
\cite{haho} who investigated the continuum limit of the $SO(4)$ matrix model. 
The measured ratio of the mass gap to $\L_\MS$ was compared with the theoretical 
prediction obtained using the Bethe ansatz \cite{holl}. There was a disagreement between 
theory and experiment by about a factor of four the measured correlation length
being about four times smaller than expected. However, the measurement using the covering 
group was in excellent agreement with theory. The numerical method used, due to 
L\"uscher et al. \cite{luea}, relies in part on measuring the correlation length in a 
large volume and establishing that scaling holds with only small and perturbative violations. 
Although in the $SO(4)$ case there were strong indications that the results scaled, the 
discrepancy between simulation and theory led to the conclusion that the signal for scaling 
was only apparent and that a true continuum limit had not been achieved in the large 
volume simulation. It was conjectured that the cause of the deception was 
the presence of vortices in the $SO(4)$ model which are absent in the case of the
covering group since
\ben
\Pi_1(SO(4))~=~Z_2~,~~~~~~\Pi_1(SU(2))~=~0~.
\een
One question is, therefore, whether a bogus signal for scaling can be observed in
the presence of vortices in two dimensions. In the work presented here this 
question is addressed in the context of an $RP^2$ gauge theory which allows 
an interpolation between the pure $RP^2$ and $O(3)$ spin models. This
gauge model contains $Z_2$ vortices coupled to a chemical potential.
We observe the conventional $O(3)$ renormalized trajectory and show that
our results are consistent with the known three-loop $\gb$ function.
We establish the existence of a first-order transition, first suggested by 
Solomon et al. \cite{SSD}, for which the order parameter is the vorticity. The
critical point terminating this first-order line will be in the domain of
a new `vorticity' fixed point. Using Monte-Carlo renormalization group (MCRG) 
techniques we observe certain flows on which the blocked observables scale and
suggest that these scaling flows are due to the influence of a nearby renormalized 
trajectory which gives rise to the possibility of the existence of a fixed point
other than the $O(3)$ one. We argue that it is unlikely that any new fixed point
can be identified with the inferred `vorticity' fixed point. Our results strongly 
indicate that the apparent or `pseudo' scaling behaviour is due to a cross-over effect 
associated with the proximity of the new scaling flows to the line of $RP^2$ spin models 
in coupling constant space. The cross-over is between regions of high and low vorticity 
which emphasizes the crucial r\^ole of vorticity in the observed properties of the model. 
Where relevant our results confirm or complement those obtained by Solomon et al. 
\cite{SSD} in an earlier study of this model. 

Another reason for studying the $RP^2$ gauge models is that it has been 
conjectured \cite{Sokal} that in 2D the continuum limit in the $RP^2$ spin model
is distinct from that in the $O(3)$ spin model. Niedermayer et al. \cite{Nieder}
and Hasenbusch \cite{Has} have suggested that this conjecture is incorrect and
that there does exist a continuum limit in the $RP^2$ model which is controlled
by the $O(3)$ fixed point. The essential question is whether or not the $RP^2$ model
is in the same universality class as the $O(3)$ model. By using MCRG methods
to show the topology of renormalization group trajectories in the $RP^2$ gauge
theory we find that a consistent and simple interpretation of our results 
is that the $RP^2$ and $O(3)$ models are in the same universality class: an
interpretation which supports the conclusions of Niedermayer et al. \cite{Nieder} and 
Hasenbusch \cite{Has}.

All results are for $RP^2$ gauge models but the simulation can be generalized
to $RP^{N-1}$ and a cursory investigation for $N > 3$ has indicated that broadly similar
results hold for this general case.

In section \ref{model} we define the model under study; in section \ref{sim} we briefly 
describe the simulation techniques; in section \ref{mcrg} we define the Monte-Carlo 
renormalization group method used and describe the measurement procedure; in section 
\ref{results} we present the results and discussion and in section \ref{conclude} we 
draw our conclusions.

\newsection{\label{model}The model}
The action used is
\[
S(\{\bS\},\{\bs\})~=~-\gb\bb \sum_{\sbx,\sbmu}\,\bS_\sbx\cdot\bS_{\sbx+\sbmu}\,\s_{\sbx,\,\sbmu}~+
                    ~\mu\,\sum_\sbx\,P_\sbx(\s) \eb~,\label{ACTION}
\]
where $\bx = (x_1,x_2)$~, $x_1,x_2 \in Z,~1 \le x_1,x_2 \le L$, labels the 
sites of an integer 2D square lattice of side $L$, and $\bmu$ takes values in $\bmu_1
= (0,1),~\bmu_2 = (1,0)~$. The spin $\bS_\sbx$ is a unit length three-component vector 
at site $\bx$ and $\s_{\sbx,\,\sbmu}$ is a gauge field on the link $(\bx,\bmu)$ taking
values in $[1,-1]$~. The plaquette of gauge fields is denoted by $P_\sbx(\s)$~ where
\ben
P_\sbx(\s)~=~\s_{\sbx,\,\sbmu_1}\,\s_{\sbx+\sbmu_1,\,\sbmu_2}
              \,\s_{\sbx+\sbmu_2,\,\sbmu_1}\,\s_{\sbx,\,\sbmu_2}~.
\een
This action is invariant under the gauge transformation
\bea
\bS_\sbx~&\rightarrow&~g_\sbx\,\bS_\sbx~, \nn\\
\s_{\sbx,\,\sbmu}~&\rightarrow&~g_\sbx\,\s_{\sbx,\,\sbmu}\,g_{\sbx+\sbmu}~,
\eea
with $g_\sbx \in [1,-1]~$.

Vortices reside on plaquettes where $P_\sbx(\s)=-1$, and are suppressed (enhanced)
if the chemical potential, $\mu$, is positive (negative). The pure $O(3)$ model
corresponds to $\mu \rightarrow \infty$ and the pure $RP^2$ model corresponds
to $\mu=0$~.  

\newsection{\label{sim}The simulation}

A local update was used comprising a combination of heat-bath, microcanonical
and demon schemes. For fixed gauge fields the spins $\{\bS\}$ were first updated by a 
heat-bath algorithm which can be generalized to $O(N)$ and so for this section we
will consider $\bS_\sbx$ to be an $N-$component spin of unit length. The heat-bath method
is to project each spin onto a 3D subspace of the $N$-dimensional space in which the 
spins take their values. The 3D subspace is chosen at random but is the same for all
spins during one lattice update. Let the projection of $\bS_\sbx$ onto this space
be denoted $\bR_\sbx$. Then clearly
\ben
(\bR_\sbx)_i~=~(\bS_\sbx)_{j_i}~,~~~~i=1,2,3~,~~~~1 \le j_1 < j_2 < j_3 \le N~,
\een
where the $j_i$ are chosen randomly subject to the restrictions above. The 
single-site probability distribution for $\bR_\sbx$ is then 
\ben
Q(\bR_\sbx)~\propto~\exp(\bM_\sbx\cdot\bR_\sbx)~,
\een
where
\ben
\bM_\sbx~=~\gb\,\sum_\sbmu\,\bb\bR_{\sbx+\sbmu}\,\s_{\sbx,\,\sbmu}~+
          ~\bR_{\sbx-\sbmu}\,\s_{\sbx-\sbmu,\,\sbmu}\eb~.
\een

The heat-bath update at each site is done by replacing $\bR_\sbx$ by $\bR_\sbx^\prime$
which is chosen with the probability distribution $Q(\bR_\sbx)$. The microcanonical spin 
update is 
\ben
\bS_\sbx~\rightarrow~\bS^\prime_\sbx~=
        ~-\bS_\sbx~+~{2\,(\bS_\sbx\cdot\bM_\sbx)\,\bM_\sbx \over |\bM_\sbx|^2}~.
\een
The demon update is applied to the gauge fields only. In general, it is only necessary
to introduce one demon variable for the whole gauge configuration.
However, when running on a massively parallel computer it is necessary
to have one demon per processor and then each demon must migrate through 
the whole lattice. This is easily achieved by moving demons sequentially
between processors. We illustrate the method with one demon variable $d,~d \ge 0$.
The action in equation (\ref{ACTION}) is augmented by the demon to become
\ben
S_{demon}(\{\bS\},\{\s\},d)~=~S(\{\bS\},\{\s\})~+~\gb\,d~.
\een
Then for each link $(\bx,\bmu)$ the trial gauge field update is 
$(\s_{\sbx,\,\sbmu}\,,d)~\rightarrow~(-\s_{\sbx,\,\sbmu}\,,d^\prime)$
where $d^\prime$ is chosen so that $S_{demon}$ is unchanged. That is,
{
 \colsep
 \bea
 &&d^\prime~= \nn\\
 &&d - 2\,\s_{\sbx,\,\sbmu}\,\left\{\bS_\sbx\cdot\bS_{\sbx+\sbmu}+
     \mu\bb\s_{\sbx,\,\sbnu}\,\s_{\sbx+\sbnu,\,\sbmu}\,\s_{\sbx+\sbmu,\,\sbnu}+
      \s_{\sbx-\sbnu,\,\sbnu}\,\s_{\sbx-\sbnu,\,\sbmu}\,
       \s_{\sbx-\sbnu+\sbmu,\,\sbnu}\eb\right\}, \nn\\
 &&
 \eea
}
where $\bnu$ is the orthogonal vector to $\bmu$. The update is accepted only
if $d^\prime \ge 0$. Note that the update is microcanonical in the augmented
configuration space of fields plus demon and hence it is independent of $\gb$.

One complete lattice update consisted of one heat-bath update followed by
an alternating sequence of $N_{MD}$ microcanonical and demon updates. The 
value of $N_{MD}$ that optimizes the decorrelation of the configurations 
depends on many factors and we did not spent much effort in tuning $N_{MD}$ 
but regard $N_{MD} \approx 10$ as a reasonable value.  The heat-bath 
update took about ten times the time of the combined microcanonical 
and demon updates and so there was little time penalty for this choice.
Depending on the coupling constant values we found that decorrelated
configurations were produced within 2-30 iterations. Lattice sizes 
ranged from $64^2$ to $512^2$, and typically the numbers of configurations per run 
were e.g., $~2\cdot 10^6$ for $64^2$ and $~5\cdot 10^5$ for $256^2$.

The simulations were carried out on the HITACHI SR2201 computers in the 
Cambridge High Performance Computing Facility and in the Tokyo Computing Centre.

\newsection{\label{mcrg}The Monte-Carlo renormalization scheme}

The objective is to establish the topology of renormalization group flows
in the relevant large-scale variables and infer the phase structure of the model. 
After sufficient blocking we assume that we are dealing with renormalized observables,
and so different phases will be distinguished by singularities in the renormalization
group flows. This has been discussed, for example, by Nienhuis and Nauenberg \cite{NN} and by
Hasenfratz and Hasenfratz \cite{haha}. We assume that there are at most two relevant
couplings in the neighbourhood of any fixed point in which we are interested.
We also assume that the chosen blocked operators have components which span
the two-dimensional space of relevant operators, i.e., the operators conjugate 
to these relevant couplings. From our earlier experience \cite{haho} and
from the surmize stated in the introduction that vorticity plays a vital r\^ole,
we chose to study how the mean values of the spin-spin interaction, $A$, and
of the plaquette, $P$, flow under blocking. For a given configuration
these quantities are defined by
\bea
A~&=&~{1\over 2V}\sum_{\sbx,\sbmu}\,\bS_\sbx\cdot
      \bS_{\sbx+\sbmu}\,\s_{\sbx,\,\sbmu}~,\nn\\
P~&=&~{1\over V}\sum_\sbx\,P_\sbx(\s)~. \label{AP}
\eea
$\AE$ lies in $[0,1]$ and $\PE$ lies in $[-1,1]$ and the mean vorticity is defined by
$\V = (1-P)/2~$. 

For each configuration $\{\bS,\s\}$ on a lattice of side $L$ we derive a
blocked configuration $\{\bS^B,\s^B\}$ on a lattice of side $L/2$~. The
blocking transformation for the spins is
\bea
&&\bS^B_{\sbx_B}~=~ \nn\\
&&{\bS_\sbx~+~\a\bb\,\bS_{\sbx+\sbmu_1}\s_{\sbx,\,\sbmu_1} +
    \bS_{\sbx+\sbmu_2}\s_{\sbx,\,\sbmu_2} + 
     \bS_{\sbx-\sbmu_1}\s_{\sbx-\sbmu_1,\,\sbmu_1} + 
       \bS_{\sbx-\sbmu_2}\s_{\sbx-\sbmu_2,\,\sbmu_2}\eb
         \over \left| \mbox{numerator}\right|}~. \nn\\ \label{SBLOCK}
&&
\eea
Based on earlier work by Hasenbusch et al. \cite{goea} the parameter $\a$ was chosen 
to be 0.0625~. Choosing other reasonable values for $\a$ was found not to change any
outcome or conclusion. This gauge-invariant blocking transformation is shown in 
figure \ref{f1}.

To block the gauge field the products of gauge fields were computed for the 
three Wilson lines joining the end points of the blocked link shown in figure \ref{f1}.
These field products are denoted by $W_0, W_+, W_-$. For the blocked link 
joining $\bx$ to $\bx+2\bmu$ the $W_i$ are given by
\bea
W_0~&=&~\s_{\sbx,\,\sbmu}\s_{\sbx+\sbmu,\,\sbmu}~,\nn\\
W_+~&=&~\s_{\sbx,\,\sbnu}\,\s_{\sbx+\sbnu,\,\sbmu}\,
         \s_{\sbx+\sbmu+\sbnu,\,\sbmu}\,\s_{\sbx+2\sbmu,\,\sbnu}~,\nn\\
W_-~&=&~\s_{\sbx-\sbnu,\,\sbnu}\,\s_{\sbx-\sbnu,\,\sbmu}\,
         \s_{\sbx+\sbmu-\sbnu,\,\sbmu}\,\s_{\sbx+2\sbmu-\sbnu,\,\sbnu}~,
\eea
where $\bnu$ is the orthogonal vector to $\bmu$. The blocked gauge field was assigned 
the majority sign of the $W_i$: 
\ben
\s^B_{\sbx_B,\,\sbmu_B}~=~{W_++W_0+W_- \over \left| W_++W_0+W_- \right|}~.\label{VBLOCK}
\een
This blocking transformation has the important property that it ensures
that two vortices on adjacent plaquettes of the original lattice will cancel
and not survive in the blocked lattice. This is clearly true if the vortices lie
in the same $2\times 2$ block since they add mod$_2$, but the majority rule guarantees 
cancellation also when two adjacent vortices lie either side of the block link separating 
two neighbouring blocks. This is illustrated in figure \ref{f2}.

For a given pair of coupling constants $(\gb,\mu)$ and given lattice size $L\times L$,
each configuration was blocked by successive transformations until the
blocked lattice size was $8\times 8$. The operator expectations $\la A \ra_L(\gb,\mu)$ and 
$\la P \ra_L(\gb,\mu)$ were then measured and averaged over all configurations. For given 
$(\gb,\mu)$ this was done for $L=64,128,256,512$ which gives four points on a segment of 
a flow  in the $(\AE,\PE)$ plane with each point labelled by initial lattice size. Each point 
corresponds to a rescaling of length by a factor of two compared with the previous point.

The errors on the observables were determined by averaging the results for successive
configurations in bins of $2^M,~M=0,1,2,\ldots$ and calculating the errors on
the ensemble of bin-averaged measurements \cite{hoho}. The true error is the 
asymptotic value achieved for large enough $M$. The decorrelation length can
also be estimated from the behaviour of the error as a function of $M$. The number
of independent configurations ranged from about 600 for $L=512$ to in excess of
$2\cdot 10^4$ for $L=64$~. Errors were also estimated from the ensemble of independent
measurements from different processors.

\newsection{\label{results}Results}

Each flow segment consisting of four points, but more complete flows can be built up by extending 
the flow in either direction by tuning to new couplings $(\gb^\prime,\mu^\prime)$ so that
\bea
\la A \ra_{L^\prime}(\gb^\prime,\mu^\prime)~&=&~\la A \ra_L(\gb,\mu)~,\nn\\
\la P \ra_{L^\prime}(\gb^\prime,\mu^\prime)~&=&~\la P \ra_L(\gb,\mu)~,\label{TUNE}
\eea
for some $L$ and $L^\prime$. The flow for given $(\gb^\prime,\mu^\prime)$ can then be
computed. In general this will be an approximate procedure because a segment in the $(\AE,\PE)$ 
plane is the projection onto this plane of part of a full flow in the higher dimensional space
of observables.  By tuning as described we are ensuring only that the projections of
blocked points coincide not the blocked points themselves. In principle, we need to 
match a full complement of observables by tuning a complete set of couplings, conjugate to
these observables, which define the most general action consistent with the symmetry. 
In general, the effect of couplings which are not included cannot be properly taken into 
account. However, we assume that in the neighbourhood of a fixed point there will be at most 
two relevant couplings and that the projection onto the $(\gb,\mu)$ plane of the space they
span is non-singular. The effect of irrelevant couplings is mitigated by performing an
initial blocking by a factor which will significantly reduce the errors induced by the
projection so that we are effectively dealing with renormalized operators. Since the target 
lattice is always $8\times 8$ the size of this factor depends on the initial lattice size 
and hence on available CPU. For our study this initial blocking factor had a minimum 
value of eight. 

There are errors due to finite-size effects which can be parametrized in terms of the
parameter $z=\xi_L/L$ where $\xi_L$ is the correlation length on the $L\times L$ 
lattice. Because the target lattice is the same size throughout, the values of $z$ 
associated with coinciding points, equation (\ref{TUNE}), are similar and so  the mismatch in 
the finite-size errors between different segments joining up to make a longer flow will be 
minimized. There will nevertheless be a residual finite-size effect which it is generally not 
possible to estimate except in the case of the $O(3)$ spin model which is discussed in the next 
section.

The details of the flows can depend on $z$ and the details of the blocking scheme. It follows 
that conclusions about the physical properties of the theory can be deduced only from universal 
or topological properties of the flows such as the occurrence of fixed points and singular 
behaviour.

\newsubsection{\label{o3}The {\boldmath $O(3)$} renormalized trajectory}

In the limit $\mu \rightarrow \infty$ we recover the $O(3)$ spin model, and to test 
our procedures and assumptions we should, at the very least, be able to
recover the perturbative $\gb$-function for this model. The projection of the $O(3)$ 
renormalized trajectory onto the $(\gb,\mu)$ plane is the $\gb$-axis. We expect
corrections to scaling due to finite lattice-spacing artifacts which will be a function
of $L$. We find that our method works well for sufficiently large $\gb$ once the tree-level 
approximation for these scaling corrections has been taken into account. Consider the block 
observable $A^B = \bS^B_{\sbx_B}\cdot\bS^B_{\sby_B}$ with the blocking transformation 
defined in equation \ref{SBLOCK}. For large $\gb$ we write
\ben
\bS_\sbx~=~\bq \sqrt{1 - \gb^{-1} \bphi_\sbx^2}~+~\gb^{-1/2}\bphi_\sbx~,
\een
where $\bq = (1,0,0)$ and $\bphi = (0,\phi_1, \phi_2)$. Using \ref{SBLOCK} and keeping
terms up to $\gb^{-1}$ we find after one blocking step that
\ben
\bS^B_{\sbx_B}~=~\bq \sqrt{1 - \gb^{-1} \bPhi^2_{\sbx_B}}~+~\gb^{-1/2}\bPhi_{\sbx_B}~,
\een
with
\ben
\bPhi_{\sbx_B}~=~{\bphi_\sbx + \a\sum_\sbmu \bphi_{\sbx+\sbmu} \over 1 + 4\a}~,
\label{PBLOCK}
\een
and where $\bmu$ is summed over nearest neighbour links. For large $\gb$ we expect
the blocked expectation value, $\la A \ra_L$, on the $8\times 8$ lattice to behave as
\ben
\la A \ra_L~=~1~-~C(L)/\gb~+~O({1/\gb^2})~.\label{CDEF}
\een
For large enough $L$ we expect $C(L)$ to attain its limiting value. However, there
is still some variation in $C(L)$ for the values of $L$ we are using. In order to
accommodate the bulk of this correction to scaling we define the effective coupling
$\gb_{eff}$ by
\ben
\gb_{eff}(L,\gb)~=~{C(L) \over (1-\la A \ra_L(\gb))}~,\label{BEEF}
\een
and modify the matching condition of equation (\ref{TUNE}) in this case to become
\ben
\gb_{eff}(L,\gb)~=~\gb_{eff}(L^\prime,\gb^\prime)~.\label{TUNE_O3}
\een
We then expect that
\ben
\log(L/L^\prime)~=~\int_u^{u^\prime}\,{du \over \gb(u)}~,\label{BETAFN}
\een
where $u = 1/\gb,~u^\prime = 1/\gb^\prime$. 

$C(L)$ is determined from a free field theory calculation on an $L^2$ lattice
of the kinetic term for the blocked field $\bPhi_\sbx$ which is defined on the 
target $L_B^2$ lattice by iteration of equation (\ref{PBLOCK}). This calculation is 
done easily numerically and the results are given in table \ref{t0}. We use $L_B=8$ 
in subsequent calculations.

\btab[htb]                                                                        
\bec                                                                             
\btabu{|c|c|c|}\hline                                                   
$L$&$C(L,L_B=8)$&$C(L,L_B=4)$\\\hline
16&0.53890071&0.54363018\\\hline
32&0.56795927&0.55935100\\\hline
64&0.58370587&0.56866918\\\hline
128&0.59303041&0.57453025\\\hline
256&0.59889307&0.57829938\\\hline
512&0.60266260&0.58074276\\\hline
\etabu
\enc
\caption[]{\label{t0}\small The function $C(L)$ defined in equation (\ref{CDEF}) for 
blocking from an $L^2$ lattice for target lattices with $L_B = 8,4~$. $C(L)$ gives the
tree approximation for the dependence on $L/a$ of corrections to scaling in the $O(3)$
spin model.}
\etab

Using equation (\ref{BETAFN}) and the three-loop $\gb$-function from \cite{luea} we
determine sequences for the bare coupling, $\gb$, for which successive terms correspond
to blocking by a factor of two. In table \ref{t00} we compare the values of $\gb_{eff}$
for the two sequences $\gb = 5.0,\,4.8861,\,4.7721$ and $2.0,\,1.8803,\,1.7560$. If
our simulation reproduces the correct $\gb$-function then the matching condition 
\ben
\gb_{eff}(2^{(m-n)}L,\gb_n)~=~\gb_{eff}(L,\gb_m)~,\label{MATCH}
\een
must be satisfied, where $\gb_n$ is the $n$-th term in the sequence. From table \ref{t00}
we see that this condition is indeed very well satisfied for the sequence starting with
$\gb = 5.0$. For the other sequence the larger values of $\gb_{eff}$ agree well and
only as $\gb_{eff}$ decreases is there an increasing discrepancy which signals a significant deviation 
of the three loop approximation to the $\gb$--function from the correct value and also the possible effect 
of the neglected $L$-dependent $O(1/\gb^2)$ terms in equation (\ref{BEEF}). Our expectation is confirmed 
that the method correctly reproduces asymptotic scaling and probes the renormalization group flow close 
to the renormalized trajectory.

\btab[htb]
\bec   
\btabu{|c||c|c|c|c|}\hline
&\multicolumn{4}{|c|}{Initial lattice size, $L$}\\\cline{2-5}
\rb{$\gb$}&64&128&256&512\\\hline\hline
5.0&--&4.355(1)&4.227(3)&4.107(5)\\\hline
4.8861&4.343(1)&4.244(3)&4.110(3)&3.995(5)\\\hline
4.7721&4.249(1)&4.124(1)&4.002(3)&--\\\hline\hline
2.0&--&1.2752(4)&1.1366(4)&0.9852(7)\\\hline
1.8803&1.2733(2)&1.1289(4)&0.9729(7)&0.8194(4)\\\hline
1.7560&1.1210(3)&0.9532(4)&0.8000(2)&--\\\hline
\etabu
\enc
\caption[]{\label{t00}\small Values of $\gb_{eff}(L,\gb)$ for sequences of bare coupling
$\gb$ computed using equation (\ref{BETAFN}) where successive couplings in the sequence
correspond to blocking by a factor of two. The required matching condition, equation 
(\ref{MATCH}), for verification of asymptotic scaling is very well satisfied for the
sequence at larger $\gb$ and the deviation in the other sequence is largest for the smaller
values of $\gb$ and is due to deviation of the three--loop approximation to
the $\gb$--function from the true value and possibly to terms neglected in equation (\ref{BEEF}) .}
\etab

\newsubsection{\label{scaling}New scaling flows}
In figures \ref{f3} to \ref{f6} we plot the flow segments for various $(\gb,\mu)$ 
values in the $(\AE,\PE)$ plane, where the longer flows in figures \ref{f3} and
\ref{f4} are composed of superimposing segments using equation (\ref{TUNE}).
There is a flow on which the observables scale. This is shown in 
figures \ref{f3} and \ref{f4}. Nearby flows also showed scaling but are 
not included in the figures for reasons of clarity. To see that observables 
scale each flow of four points was successively
overlaid using the tuning described in equation (\ref{TUNE}) with $L^\prime = L/2$.
For the scaling flows the points of the overlaid flow segments coincide very well within
errors (the tuning is not absolutely exact) showing that there are only very small
effects from the irrelevant operators. In table \ref{t1} we give the values 
of $(\AE,\PE)$ for points on the segments making up
two such scaling flows, labelled set 1 and set 2, together with the initial lattice 
sizes and coupling constant values. 

The tuning described by equation (\ref{TUNE}) was done by trial and error. We
tried to estimate the effect of small increments in the coupling constant values
by using the method of reweighting but this was found not to work. This was mainly
because the blocked $(\AE,\PE)$ values were very sensitive to the initial couplings and
so to use reweighting was not a realistic possibility. Also it was found that the effect
of a change in $\gb$ could be partly compensated by a change in $\mu$. This was because
the Jacobian $\d(\AE,\PE)/\d(\gb,\mu)$ was relatively small and presumably a better choice
for the pair of observables and/or initial couplings would increase its value. However,
in no case was this Jacobian dangerously small and tuning by trial was effective.

For these flows we did not apply a compensation for scaling corrections of the kind
used in the previous section for the $O(3)$ model. It is not possible to carry out
a similar tree-level calculation to determine the compensation, if any, as a function
of $L/a$ since the model and the appropriate coupling(s) associated with the scaling
flows are not known. Also, the data presented in this section were obtained
before the details of the $O(3)$ analysis were known and the possible residual dependency 
on $L$ analyzed. Overall, the deviation from scaling for the scaling flows is not large and
we must allow for the possibility that it is due, in some measure, to the finite lattice-spacing 
artifact of the kind analyzed for $O(3)$. However, there are a few points shown which do 
deviate substantially from scaling for which such an explanation is unlikely. For both sets 
these points are for the two largest values used for $\gb$, and for the largest initial 
lattice size, $L=512$. We defer the discussion of the reasons why they do not scale well 
until section {\ref{discussion}}.

\btab[htb]
\bec
\btabu{|c|c||c|c|c|c|}\hline
&&\multicolumn{4}{|c|}{Initial lattice size, $L$}\\\cline{3-6}
&\rb{$(\gb,\mu)$}&64&128&256&512\\\hline\hline
&&0.81894(3)&0.80587(9)&0.7913(2)&0.7761(3)\\
&\rb{4.26, -0.045}&0.98330(4)&0.9809(2)&0.9753(3)&0.9684(8)\\\cline{2-6}
&&0.80793(3)&0.7913(2)&0.7690(2)&0.739(1)\\
&\rb{4.0, 0.0}&0.98021(6)&0.9740(3)&0.9561(4)&0.927(3)\\\cline{2-6}
&&0.79261(5)&0.7678(1)&0.7273(4)&0.6424(8)\\
&\rb{3.72, 0.06}&0.97402(9)&0.9547(2)&0.9053(8)&0.770(3)\\\cline{2-6}
&&0.76947(6)&0.7251(2)&0.6363(4)&0.4995(6)\\
&\rb{3.45, 0.125}&0.9576(1)&0.9025(4)&0.7601(10)&0.558(1)\\\cline{2-6}       
&&0.7233(2)&0.6298(2)&0.4872(2)&0.2878(5)\\
\raisebox{20mm}[0pt]{Set 1}&\rb{3.18, 0.187}&0.9025(4)&0.7528(4)&0.5430(4)&0.3348(5)\\\hline\hline  
&&0.81574(3)&0.80228(6)&0.7871(2)&0.7694(9)\\                           
&\rb{4.00, 0.02}&0.98787(5)&0.9849(1)&0.9782(4)&0.967(2)\\\cline{2-6}       
&&0.80236(4)&0.7846(1)&0.7611(2)&0.720(1)\\                                   
&\rb{3.73, 0.0833}&0.98518(6)&0.9768(2)&0.9571(5)&0.905(3)\\\cline{2-6}                
&&0.78475(7)&0.7577(1)&0.7085(3)&0.611(2)\\
&\rb{3.495, 0.14}&0.9769(1)&0.9538(3)&0.8865(7)&0.727(3)\\\cline{2-6}
&&0.7596(1)&0.7070(1)&0.6062(4)&0.4564(7)\\                                  
&\rb{3.3, 0.18}&0.9544(2)&0.8826(3)&0.7178(8)&0.506(1)\\\cline{2-6}                       
&&0.7050(3)&0.6002(30&0.448(5)&0.2383(2)\\                                  
\raisebox{20mm}[0pt]{Set 2}&\rb{3.06, 0.23}&0.8821(7)&0.7105(7)&0.4990(9)&0.2973(4)\\\hline  
\etabu                                                                          
\enc
\caption[]{\label{t1}\small Values of the blocked observables $(\AE,\PE)$ for 
the flow segments constituting two neighbouring scaling flows labelled 
by set 1 and set 2. The values of the couplings $(\gb, \mu)$ labelling each segment is 
given and the different points on a given segment are labelled by the initial lattice 
size $L$. The blocked observables were measured after blocking to a fixed target lattice 
size of $8\times 8$. Scaling of the blocked observables can be seen to hold extremely
well for each set by noting that the $(\AE,\PE)$ values lying on any given diagonal sloping
from bottom left to top right agree very closely indeed, except for the largest two values
of $\gb$ for $L=512$. Set 1 is shown in figures \ref{f3} and \ref{f4}.}
\etab

For the couplings which generate the scaling flow denoted by set 1, a plot of $\exp(-\mu)$ 
versus $g = 1/\gb$ can be approximated by a straight line with all except the one with smallest
$\gb$ well fitted by
\ben
\exp(-\mu)~=~1.74(1)~-~2.98(4)g~,\label{SCT}
\een
Phenomenologically, we find that it is consistent to associate a scaling exponent, $\k$, with the 
scaling flow using the relation
\ben
\k~=~{p\log(2)/\log((g_1-g^*)/(g_2-g^*))}~.\label{KAP}
\een
where the number of blockings for $g_1$ and $g_2$ differ by $p$ (a factor of $2^p$ in
change of scale) in order for the corresponding points in the $(\la A,\ra,\la P,\ra)$ plane to coincide 
on the scaling flow. The value of $g^*$ is a free parameter which is chosen to 
obtain the best fit assuming $\k$ in equation (\ref{KAP}) is constant. Even so, $g^*$
is not very well determined by this alone and we find a range of values $g^* \approx .13 \sim .15$
to be acceptable, for which the various pairings of couplings for set 1 from table \ref{t1}
give the values for $\k$ shown in table \ref{t2}. The values of $\mu^*$ in table \ref{t2} are
inferred using the linear fit of equation (\ref{SCT}). Note that the possible values of $\mu^*$ 
in table \ref{t2} are far removed from the value $\mu = 0$ which characterizes the $RP^2$ model.
Taking $\k$ in the range $4 \sim 5$, the dimension of the relevant operator associated with 
the scaling flows is
\ben
\Delta~=~D - {1/\k}~\approx 1.75 \sim 1.8~.
\een
A word of caution. The definition of $\k$ used in equation (\ref{KAP}) is appropriate for 
conventional second-order behaviour. However, the consistency of the fit should be taken only 
as a phenomenological parametrization and it is likely that a different scaling form such as derived 
from Kosterlitz-Thouless behaviour would give an equally good fit. For example, in \cite{siea} 
Seiler et al. study the $~Z(10)$ model in 2D which they presume has a KT transition. They find
that both second-order and KT scaling forms give equally good fits near the transition and 
that, indeed, the KT form is hard to reconcile with conventional theory. 

\btab[htb]
\bec
\btabu{|c||c|c|c|c|c|c|c|c|c|c|}\hline
$(g^*,\,\mu^*)$&\multicolumn{10}{|c|}{Scaling exponent $\k$}\\\hline
0.13,\,-0.305&5.10&4.92&4.919&4.90&4.76&4.83&4.84&4.91&4.88&4.84\\\hline
0.15,\,-0.261&4.19&4.10&4.15&4.18&4.02&4.13&4.18&4.25&4.26&4.28\\\hline
\etabu
\enc
\caption[]{\label{t2}\small The scaling exponent $\k$ calculated using equation (\ref{KAP}) and
various pairings of couplings $g = 1/\gb$ from set 1 given in table \ref{t1}. Two different
choices for $(g^*,\,\mu^*)$ are used which correspond to the range giving consistent 
results for $\k$.}
\etab

\newsubsection{\label{first-order} The first-order transition}
Solomon et al. \cite{SSD} use simple arguments to suggest that a line of first-order
transitions occur in the range $\mu_1 < \mu(\gb) < \mu_2,~~\mu_1 = -0.293, \mu_2 \approx -0.26$. 
This line of transitions will terminate in a critical point associated with a continuous
transition implying that a critical surface intersects the $(\gb,\mu)$ coupling
constant plane. Confirmation of the first-order transition is shown in figures \ref{f7}
and \ref{f8} where, respectively, the values of $\AE$ and $\PE$ for a lattice of $L=128$ are shown 
plotted against $\mu$ for fixed $\gb = 6.0, 7.0, 7.5, 8.0$. It is clear that there is no 
transition for $\gb=6.0, 7.0$ but that there is likely to be a first-order transition for
$\gb= 7.5, 8.0$. This places the critical point in the range $7.0 < \gb^* < 7.5$ which is
consistent with the investigation of Solomon et al. \cite{SSD}. The value of $\mu^*$ 
varies a little but is close to $-0.26$. There is no detectable dependence on $L$ and
studies on lattices with $L=256,512$ have shown identical results within statistical errors. From
figures \ref{f7} we infer that a good order-parameter distinguishing the two phases
is the vorticity. There is also a discontinuity in $\AE$ shown in figure \ref{f8}
which is, however, not independent of the discontinuity in $\PE$. The value of $\AE$ will vary 
rapidly since it is sensitive to, and thus reflects, the discontinuous change in the vorticity.
The effective potential will show no discontinuity and it is crude but reasonable
to suppose that one particular linear combination of $A$ and $P$ plays this r\^ole.
In figure \ref{f9} we plot $\PE$ against $\AE$ and indeed it can be seen that there is
no sign of a sharp discontinuity and that the locus of points is reasonably linear. The
outcome is that the combined operator
\ben
U_C(\bx)~=~\sum_\sbmu\,\bS_\sbx\cdot\bS_{\sbx+\sbmu}\,\s_{\sbx,\,\sbmu}~+
          ~\g P_\sbx(\s)~,\label{UC}
\een
with $\g \approx -0.29$, has a continuous expectation value across the transition. The 
orthogonal combination, $U_D(\bx)$, has a discontinuous expectation value which is 
sensitive to the vorticity and is a good order-parameter. The simple and persuasive argument 
of Solomon et al. \cite{SSD} is based on minimizing the energy at $\gb=\infty$ to show that a first-order 
transition occurs in $\PE$ at $\mu(\infty) = 1/\sqrt{2}-1$. The argument also depends on continuity 
of the energy which, at $\gb=\infty$, means that $U_C(\bx)$ is identified with the local energy operator. 
This will be only approximate for $\gb < \infty$.

The critical point terminating the first-order line will be in the domain of a fixed point,
the `vorticity' fixed point, with a renormalized trajectory on which the vortex density scales 
thus defining a new continuum theory with non-zero vortex density.

\newsubsection{\label{crossover}Cross-over of flows}
In figures \ref{f5} and \ref{f6} there are two further sets of flow segments, sets 3 and 4,
which each show a clear cross-over as a function of initial couplings. Set 3 lies to the
left of set 4. These two sets are examples of the cross-over effect which we infer
occurs in a narrow region formed by the neighbourhood of a continuous line of theories 
in the $(\gb,\,\mu)$ plane. 

The couplings associated with sets 3 and 4 are listed in table \ref{t3} where, for
each set, the couplings reading from left to right label the segments in
order from the lowest in the figure (lowest $\PE$) to the highest (highest $\PE$).

\btab[htb]
\bec
\btabu{|c||c|c|c|c|c|c|c|c|}\cline{2-7}
\multicolumn{1}{c}{}&\multicolumn{6}{|c|}{$(\gb,\,\mu)$}\\\hline
&3.3&3.3&3.3&3.3&3.3&3.3\\
\rb{Set 3}&0.34&0.36&0.40&0.45&0.50&0.60\\\hline
&3.9&4.05&4.17&4.19&4.29&4.5\\
\rb{Set 4}&0.0&0.0&0.0&0.0&0.0&0.0\\\hline
\etabu
\enc
\caption[]{\label{t3}\small The couplings associated with sets 3 and 4. For each set,
the couplings reading from left to right label the segments in
order from the lowest in the figure (lowest $\PE$) to the highest (highest $\PE$).}
\etab

We concentrate in particular on the flow segments of set 4 which correspond to 
the $RP^2$ spin model ($\mu = 0$). These segments are shown for various $\gb$ in the  
range $3.9~-~4.5$ where the cross-over in the flows is very strongly marked, occurring between 
$\gb=4.17$ and $\gb=4.19$. The values of $(\la A,\ra,\la P,\ra)$ blocked from $L=64 \rightarrow L=8$
show little variation in values but when blocked from larger $L$ the variation is very
strong indeed. Moreover, it is clear that for $\gb = 3.9 \sim 4.1$ the flow is dominated
by the proximity of the scaling flows, represented as the dashed line in the figure, 
which will induce a signal for scaling in pure $RP^2$ when $\gb \sim 4$. However, as $\gb$ is 
increased there is a strong crossover effect, and for $\gb > 4.25$ the flow is consistent with 
dominance by the renormalized trajectory associated with the asymptotically free $O(3)$ fixed point at 
$(\gb^*,\,\mu^*) = (\infty, \infty)$. Consequently, the apparent scaling signal will only be 
transitory. For finite $\gb$ there will always be some free vortices, but for
$\gb > \gb_{cross-over}$ we expect that the density of free vortices will 
vanish faster than $1/\xi^2$ and the vortex density will not scale. This behaviour is consistent
with the absence of a phase transition at finite $\gb$ as well as with the $O(3)$ type
continuum limit in $RP^2$.

\newsection{\label{discussion}Discussion}
The observation of the scaling flows reported in section \ref{scaling} poses the question
of whether we can attribute them to the influence of a nearby renormalized trajectory 
and so infer the existence of a new fixed point. One interpretation of the evidence for scaling is that 
a new renormalized trajectory exists with exponent $\k \approx 4 \sim 5$, and that a new fixed point lies 
somewhere in the complete space of coupling constants with projection onto the $(\gb,\mu)$ plane of 
$(\gb^* \approx 7, \mu* \approx -0.28)$. Evidence presented in section \ref{first-order}
shows that there is also a `vorticity' fixed point associated with the `vorticity' critical point 
located at about $(\gb_c \approx 7, \mu_c \approx -0.26)$ which terminates the first-order line. Because 
of the proximity of these two points to each other it is tempting to identify the new fixed point
with the `vorticity' fixed point. However, there is no direct evidence
that this is so and there are arguments against such an identification. The first is that
we would expect the continuum limit defined at the `vorticity' fixed point to be ising-like
since the order parameter is based on a locally discrete variable: the plaquette operator, $P_\sbx(\s)$.
The exponent $\k$ for ising-like critical points is $\k = 8/15$ whereas for the new renormalized
trajectory we find $\k \approx 4 \sim 5$. This is clearly inconsistent with the proposal. The
second argument is that the identification of the two fixed points means that the `vorticity' critical 
point is in the domain of attraction of the new fixed point. The consequence is that there must be 
a fixed point of the flows in the $(\AE,\PE)$ plane in the limit that the initial lattice size 
is large enough: $L \rightarrow \infty$. This is true because flows that have bare couplings held
at the critical point values must be in the critical surface and so flow towards the new fixed point.
In turn this implies that the correlation length $\xi_A(\gb,\mu)$ for a state interpolated by $A$ must 
diverge at the `vorticity' critical point, $(\gb_c,\mu_c)$. This is unlikely since we expect 
$\xi_A(\gb,\mu)$ to be bounded from above by the correlation length in the O(3) model at the
same $\gb$, namely $\xi_A(\gb,\infty)$. This is because for $\mu < \infty$ the presence of vortices 
introduces disorder in the system which acts to reduce the correlation length at fixed $\gb$.
However, $\xi_A(\gb,\infty)$ diverges only in the limit $\gb \rightarrow 0$, the $O(3)$ fixed point,
and hence $\xi_A(\gb,\mu)$ cannot diverge at $(\gb_c,\mu_c)$ contradicting the proposed identification 
of the two fixed points. Although we did not carry out an exhaustive investigation we found no evidence 
for a fixed point of the $(\AE,\PE)$ flows from the simulations described in section \ref{scaling}. 
  
In figure \ref{f10} we shown an artist's impression of a possible topology of the RG flows in coupling
constant space consistent with this interpretation and with the results presented in 
section \ref{results}. The two axes associated with the couplings $(\gb,\,\mu)$ are augmented by a 
third which represents all other couplings. There are two fixed points shown. 
One is the usual asymptotically-free $O(3)$ fixed point and the other is the 
new fixed point we have identified in this work. The `vorticity' fixed point is not shown. The
critical surface bounds the surface of first-order transitions and the two phases associated
with this transition are distinguished by the vortex density being large in one phase and small
in the other. The line of intersection of the first-order surface with the $(\gb,\,\mu)$ plane
is the line of first-order transitions reported above. There are a number of possible continuum 
limits in this model, each identified with a different fixed point. A non-zero vortex density will 
be associated with the continuum limit taken at the critical point controlled by the `vorticity' 
fixed point. At the new fixed point there are two relevant directions but we cannot be sure what 
the relevant observables are since in this scenario the action must be augmented by other couplings 
so that it can be tuned to lie in the critical surface and in the domain of attraction of this fixed 
point. However, the presence of this renormalized trajectory dominates all flows in its neighbourhood, 
and its influence will only be diminished if points in the critical surface are approached which are
not in its domain of attraction. The example scenario of figure \ref{f10} is complicated but
we have found no simpler toplogy consistent with the results if we demand that the scaling flows
are due to a nearby RT in an extended model.

A different interpretation is that the scaling flows are due to the ghost of the Kosterlitz-Thouless
renormalized trajectory in the equivalent $O(2)$, or $XY$, model. A cogent argument against a 
Kosterlitz-Thouless fixed point occurring in non-abelian models has been given by Hasenbusch in 
\cite{Has}, but it was conjectured in \cite{haho} that some remnant of Kosterlitz-Thouless 
behaviour might nevertheless survive in models of this kind and give rise to the pseudo-scaling 
behaviour reported in \cite{haho}. As remarking in section \ref{scaling} the fit to the exponent
$\k$ using equation (\ref{KAP}) should not be taken to rule out KT behaviour in favour of conventional
second-order behaviour. Indeed, the large value for $\k$ mitigates in favour of a KT interpretation
\cite{jana}. This explanation has the virtue of simplicity over the alternative picture above but 
it is unclear how to describe the mechanism more fully.
 
The deviation from scaling for the large $\gb$ points for $L=512$ in sets 1 and 2 (table \ref{t1}) 
can be explained by noting that there is no fixed point for the $(\AE,\PE)$ flows and so the
attempt to follow the scaling flow to larger $\gb$ and into a fixed point will fail as the critical 
surface is approached. The conjectured renormalized trajectory dominates by virtue of its 
large exponent but scaling will eventually be violated as $\gb$ increases towards $\gb \sim 7$.  

The strong influence of the scaling flows gives rise to a cross-over effect in the flows which signals
the cross-over from the vortex to the spin-wave regions of the phase diagram. For example, in pure $RP^2$ this
occurs at about $\gb = 4.18,~\mu=0$.  We would naturally associate this cross-over with the 
observed first-order line but it is clear that the strength of the effect is due to the nearby 
scaling flows. This would suggest that the first-order line and the scaling flows were 
related but, as argued above, a simple relationship is ruled out and it is unclear whether the proximity 
of the two features is a coincidence or not. The region in which the cross-over occurs is quite
narrow and has been shown as a surface with dotted outline in figure \ref{f10}. As $\gb$ is increased at 
fixed $\mu$ through this `cross-over region', the vorticity rapidly decreases from a high to low value 
especially in the neighbourhood of the critical surface. This effect means that the disorder also decreases
rapidly and we would expect a corresponding rapid increase in the vector and tensor correlation lengths,
$\xi_V$ and $\xi_T$, which are deduced respectively from the correlators $G_V(\bx,\by)$ and $G_T(\bx,\by)$  
defined for $RP^{N-1}$ by
\bea
G_V(\bx,\by)~&=&~\la \bS_\sbx\cdot\bS_\sby \ra_c~, \nn\\
G_T(\bx,\by)~&=&~\la (\bS(\bx)\cdot\bS(\by))^2 \ra - 1/N~,\label{VT}
\eea
Because $G_V$ is not gauge invariant it will vanish unless it is evaluated in a fixed gauge. 
This is analogous to the situation in QED where the electron propagator is not gauge invariant but 
the pole mass is. Technically, the gauge-fixed electron propagator has a cut whose discontinuity is a 
gauge-dependent function of $\a$ but whose branch point defines the gauge-invariant mass. This is due 
to the continuous nature of the gauge group which does not apply in our case. A reasonable gauge 
choice would be to maximize $\sum_{\sbx,\sbmu} \s_{\sbx,\sbmu}\;$. $G_T$ takes the same form as 
the tensor correlator defined by Caracciolo et al. \cite{caea} and Sokal et al. \cite{Sokal} . 
Because $G_T$ is gauge-invariant it does not require gauge-fixing before evaluation. 
When the vorticity is vanishingly small the gauge field is equivalent to a pure gauge and can 
be gauge transformed to the trivial configuration $\s_{\sbx,\,\sbmu} = 1,~\forall~\bx,\,\bmu$. The 
physical observables in the theory are then insensitive to the chemical potential $\mu$ and the 
theory is in the universality class of the $O(3)$ fixed point. 

In the $O(3)$ continuum limit both $\xi_V$ and $\xi_T$ will diverge, but in the continuum limit
defined by the vorticity fixed point we expect both $\xi_V$ and $\xi_T$ to remain finite
because, as already discussed above, the presence of disorder means that they will be bounded from 
above respectively by $\xi_V(\gb,\infty)$ and $\xi_T(\gb,\infty)$, the correlation lengths at 
the same value of $\gb$ in the $O(3)$ spin model. In other words, at fixed $\gb$ we expect 
both $\xi_V$ and $\xi_T$ to increase as $\mu$ increases, achieving their maximum values at 
$\mu=\infty$ in the $O(3)$ model. This increase could be very rapid in the vicinity 
of the cross-over region. The operators interpolating the states in $G_V$ and $G_T$ are, 
respectively, $V_i = S_i$ and $T_{ij} = S_iS_j-1/N\delta_{ij}$. Since $\la V_i \ra = \la T_{ij} \ra = 0$ 
they show no discontinuity across the first-order line and hence $\xi_V$ and $\xi_T$ will not diverge
at the critical point terminating the first-order line (gauge fixing is understood where necessary).
If either of $\xi_V$ or $\xi_T$ did diverge it would contradict the expectation that they
are bounded from above by their corresponding values in $O(N)$ as mentioned above. In principle,
we could also study $G_S(\bx,\by) = \la U_C(\bx)\,U_C(\by)\ra_c$ since $\la U_C \ra$ is continuous 
across the first-order line and it couples to the S-wave two-particle $O(N)$ singlet state.
The associated correlation length $\xi_S$ should coincide with $\xi_T$ in the continuum limit
if the conventional scenario is assumed. We suggest that the divergent correlation length at 
the critical point is associated with the correlator of $U_D(\bx)$, or equivalently with 
the vorticity correlator
\ben
G_P((\bx,\by))~=~\la P_\sbx\,P_\sby\ra~.
\een
In this study $G_P$ was not computed. 

The pure $RP^2$ model ($\mu=0$) does not intersect any critical surface except the one
in the basin of attraction of the $O(3)$ fixed point at $\gb=\infty$~. This confirms the conjectures
of Niedermayer et al. \cite{Nieder} and Hasenbusch \cite{Has} that $RP^2$ and $O(3)$ have the same
continuum limit. In a simulation of pure $RP^2~$ Kunz and Zumbach \cite{kuzu} observe the rapid
decrease in vorticity that we have associated with the cross-over region, and Niedermayer et al.
\cite{Nieder} comment that in this region a sharp transition to a huge value for $\xi_V$ is to be 
expected. Our result is that the cross-over is very strongly marked in the renormalized quantities
obtained after substantial blocking has been performed. The cross-over region separates two 
phases in one of which the vorticity density is high with a background of vortices pairs overlaid by
a gas of free vortices, and in the other the vorticity density is low and does not scale as 
$\gb \rightarrow \infty$. These two phases are also separated by a first-order line and we conjecture
that a non-zero scaling limit for the vortex density could exist at the terminating critical point.
Huang and Polonyi \cite{hupo} have discussed the existence of a continuum limit with a non-zero 
scaling vorticity in a generalized 2D Sine-Gordon model and the non-conservation of the kink current. 
A similar analysis could be fruitful in non-abelian models of the kind discussed in this paper, 
although it is unclear if the same techniques are directly applicable. 

In the simulation of the 2D $SO(4)$ matrix model \cite{haho} a bogus signal for scaling was observed 
which led to an incorrect measurement of the $m/\L_\MS$ ratio. In the context of $RP^2$ we would expect
a similar effect for $\gb \sim 3.9$ because in this case the model renormalizes close to
the scaling flows and so in the neighbourhood of this coupling we should expect to 
see a good scaling signal. The effect is enhanced by the large exponent
$\k \sim 4$ associated with these flows. Simulations which are designed to compute $m/\L_\MS$
must have $\xi_V \ll L$ for some largest practical lattice size, $L$. Because $\xi_V$ is rising
rapidly in this region as a function of $\gb$, this means that only a small range of $\gb$ is
useable and that this range corresponds to theories where the vorticity is not too low since $\xi_V$
would otherwise already be too large. The conclusion is that such simulations will see an apparent
scaling due to the strong influence of the scaling flows. However, this scaling is not a signal for
a continuum limit in pure $RP^2$ but is due to the proximity of the cross-over region to the scaling
flows. On much larger lattices as $\gb$ is increased a cross-over to true scaling would
eventually be observed: the scaling associated with the $O(3)$ fixed point. However, this
would be for prohibitively large values of $\xi_V$, perhaps as large as $\xi_V \sim 10^9$
\cite{Nieder}. We believe that this effect explains the results presented in \cite{caea} 
who observe scaling in $RP^2$ ($RP^3$) but who find that the observed correlation length is
smaller by a factor of $10^7$ ($10^4$) than that deduced assuming that the theory is asymptotically 
free. We suggest that this study is actually in the cross-over regime where the correlation 
length is diminished by the disordering effect of vortices and the scaling, which is perhaps due to a new 
renormalized trajectory, is only apparent. The true scaling regime associated with the $O(3)$ ($O(4)$) 
fixed point will correspond to much larger correlation lengths than those studied. We believe 
that a similar effect caused the bogus signal for scaling in the analysis of the $SO(4)$ 
matrix model \cite{haho} and the mismatch between the observed mass-gap and the Bethe-Ansatz prediction.   

We conclude that the $RP^2$ and $O(3)$ spin models are in the same universality class and that
there is no evidence to the contrary. This confirms the conclusions of Hasenbusch \cite{Has} and
Niedermayer et al. \cite{Nieder}, but is at variance with the proposition of Caracciolo et al.
\cite{Sokal} that the continuum limits of these two models are distinct. These latter authors
propose that there is a continuous set of universality classes in a 2D model with mixed
isovector and isotensor $O(3)$ spin interactions. The $O(3)$ and $RP^2$ theories correspond to the 
pure isovector and pure isotensor interactions respectively, and the proposition of Caracciolo et al.
requires that these two models be in different universality classes. The work presented in
this paper shows that the opposite is true and hence that the existence of a continuous set
of universality classes in the mixed model is unlikely.

\newsection{\label{conclude}Conclusions}
In this paper we have studied the 2D $RP^2$ gauge model that is characterized by two couplings 
$(\gb,\mu)$, where $\mu$ is the chemical potential controlling the vorticity computed from
the gauge field plaquette expectation value. We have found that the r{\^o}le played by
the vorticity in the nature of the phase diagram is crucial. Using standard methods we
confirm the existence of a first-order transition (figures \ref{f7} to \ref{f9}), first suggested by 
Solomon et al. \cite{SSD}, in the $(\gb,\mu)$ plane separating phases of high and low vorticity. The critical point
terminating this first-order line is established to lie in the range $7.0 < \gb_c < 7.5,~\mu_c \sim -0.26$,
which implies the existence of a `vorticity' fixed point controlling the continuous transition at $(\gb_c,\mu_c)$.
We use the Monte-Carlo renormalization group for blocking the spin-spin interaction and plaquette
expectation values, $\AE$ and $\PE$, to investigate the topology of the renormalization group flows.
We verify the presence of the $O(3)$-renormalized trajectory (at $\mu = \infty$) and find results consistent 
with the known three-loop $\gb$-function for sufficiently large $\gb$ once the finite lattice-spacing artifact  
has been taken into account. We establish the existence of new scaling flows in the $(\AE,\PE)$ plane (figures 
\ref{f3} and \ref{f4}) and conjecture that they are due either to the ghost of the Kosterlitz Thouless 
renormalized trajectory in the $XY$ model or to a new renormalized trajectory and its associated fixed point which 
should lie out of the $(\gb,\mu)$ plane in the complete space of couplings. The scaling flows are consistent with a 
critical exponent $\k \approx 4 \sim 5$ and the projection of the conjectured fixed point onto the $(\gb,\mu)$ plane 
is deduced to be in the range $\gb^* \approx 6.5 \sim 7.5,~\mu^* \approx -0.31 \sim -0.26$. Although the values 
of $(\gb_c,\mu_c)$ and $(\gb^*,\mu^*)$ are very similar there are strong arguments against identifying the conjectured 
fixed point with the `vorticity' fixed point. One is that the exponent $\k$ is much larger than that expected at
the `vorticity' fixed point, and another is that such an identification would imply a fixed point
in the $(\AE,\PE)$ flows for bare couplings $(\gb_c,\mu_c)$, with a consequent divergence in certain correlation
lengths. This is contradicted by the fact that, because of the presence of non-zero vorticity, these correlation 
lengths are bounded from above by the corresponding quantities in the $O(3)$ model ($\mu = \infty$) at the 
same $\gb$ which are known not to diverge for $\gb < \infty$. A consequence is that the 
critical point at $(\gb_c,\mu_c)$ cannot be in the
domain of attraction of the conjectured fixed point. The scaling flows dominate the flows in their 
vicinity and in particular gives rise to a cross-over (figures \ref{f8} and \ref{f9}) between regions
of high vorticity (lower $\gb$) to low vorticity (higher $\gb$) associated by a rapid increase in the 
correlation length as the disorder is reduced. We conclude that simulations in the neighbourhood of 
the cross-over region for $\mu > -0.26$ will show `pseudo' scaling \cite{haho} because of the proximity 
of these scaling flows. The true continuum limit for such models will not be observed 
until true scaling, controlled by the $O(3)$ fixed point, has been established at larger $\gb$ 
and very much larger correlation length. This is the case for the $RP^2$ spin model ($\mu = 0$) 
whose continuum limit is controlled by the $O(3)$ fixed point and which is thus in the same 
universality class as $O(3)$, contradicting Caracciolo et al. \cite{Sokal} but confirming the work of 
Hasenbusch \cite{Has} and Niedermayer et al. \cite{Nieder}. It also gives an explanation for the results
discussed by Caracciolo et al. \cite{caea}. In figure \ref{f10} an artist's impression 
of the renormalization group flows is given for one scenario consistent with our results.
The natures of any new fixed points are not established because of the known difficulty \cite{siea} in
distinguishing between fits of different scaling forms and the compatibility of the observed scaling 
with a second-order scaling form, equation (\ref{KAP}) is of phenomenological significance only. 
It is quite possible that any fixed point whose existence we infer from the data is of 
Kosterlitz-Thouless type.

Our investigation has shown that the nature of gauged spin models is complicated and it is difficult to
pin down more about the nature and location of the topological features of the renormalization group flows 
without more information concerning the relevant operators in each case. However, it is clear that a fixed 
point in a larger coupling constant space can be close enough to the subspace of simple models that it very 
strongly influences observables and the outcome of tests for scaling in exactly that region accessible by simulation,
namely for those couplings for which the correlation lengths have increased to the practical limit measurable
on modern computers. This influence is strengthened if the exponent of the associated renormalized
trajectory is large. The model studied in this paper is a good example of this effect.

It would be interesting to more accurately locate the critical point at $(\gb_c,\mu_c)$ terminating the first-order
line and investigate the continuum limit it defines. 

\newsection{\label{acknowledge}Acknowledgements}
This work is supported by NATO collaborative research grant no. CRG950234 and DOE grant
number DE-FG02-85ER40237. The authors wish to thank Ian Drummond for useful conversations. 
The computing resources were provided by the High Performance Computing Centre, University 
of Cambridge and by the Computer Centre, University of Tokyo.

\newpage
\bibliography{refs}
\bibliographystyle{unsrt}
\newpage
\vfill
\bef  
\bec  
\epsfig{file=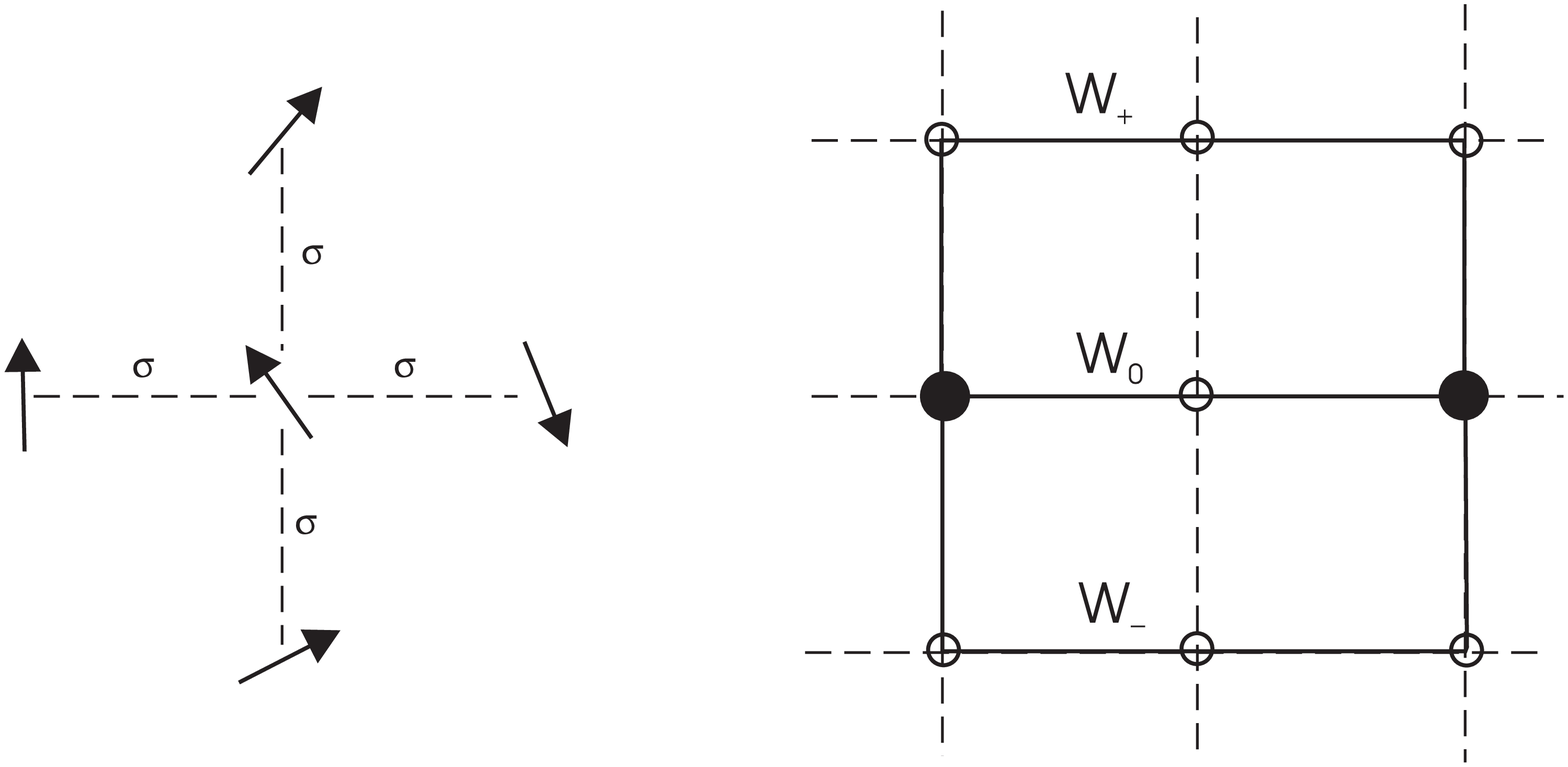,height=40mm}
\enc  
\caption{\label{f1}\small Blocking strategies for spins and gauge fields. A gauge-covariant
linear combination of a spin and its nearest neighbours defines the blocked spin and
the gauge field on the blocked link, which connects the solid-black sites, is assigned 
the majority sign of the three Wilson lines $W_+,W_0,W_-$.}
\enf
\vfill
\bef
\bec
\epsfig{file=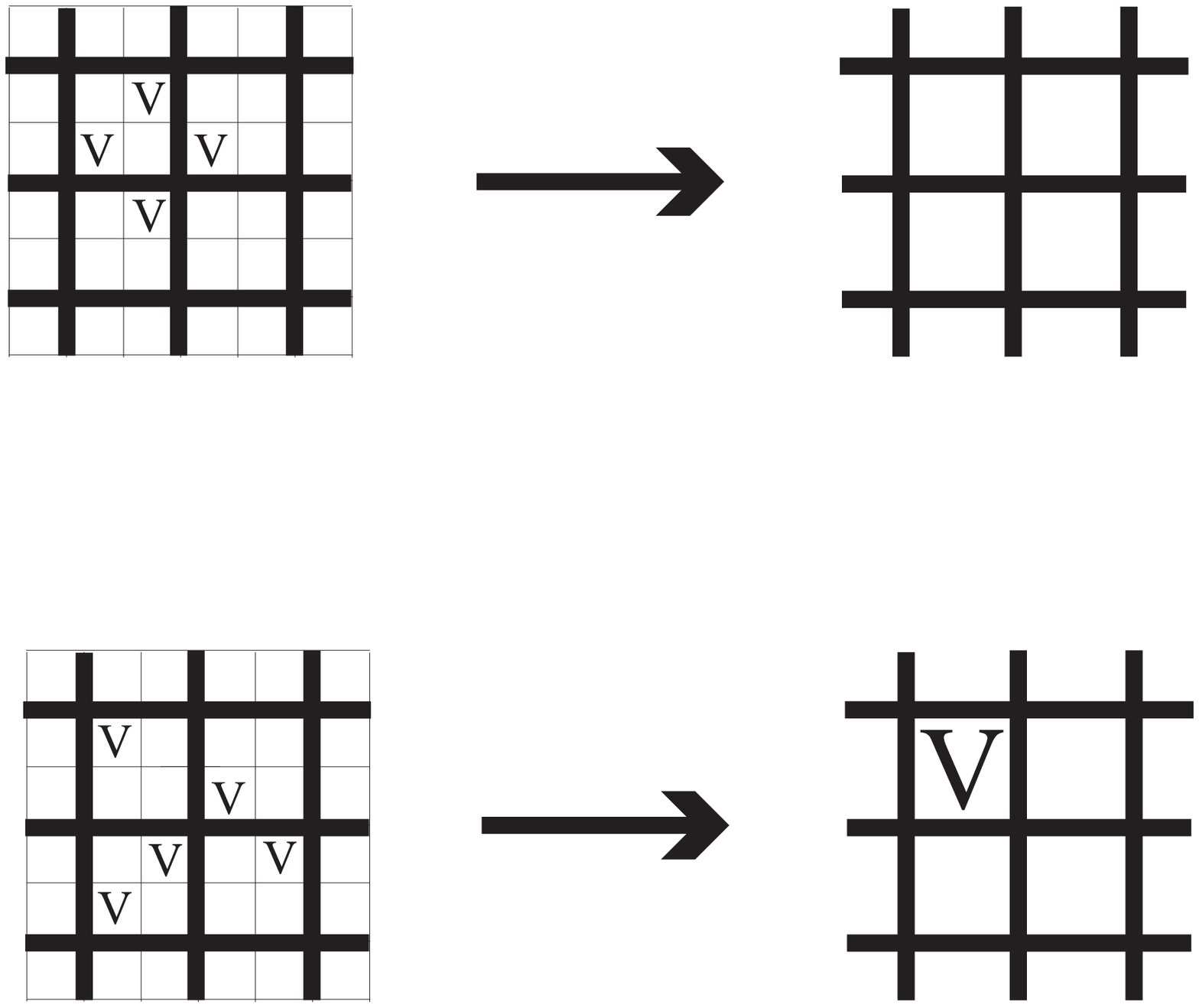,height=70mm}
\enc
\caption{\label{f2}\small The results of example blocking of two vortex configurations.
Since vortices add mod$_2$ the blocking should yield either no blocked vortices or
one blocked vortex depending on whether the original region contained an even
or odd number of vortices. The majority rule of equation (\ref{VBLOCK}0 guarantees this
important property.}
\enf
\vfill

\bef                      
\bec                                               
\epsfig{file=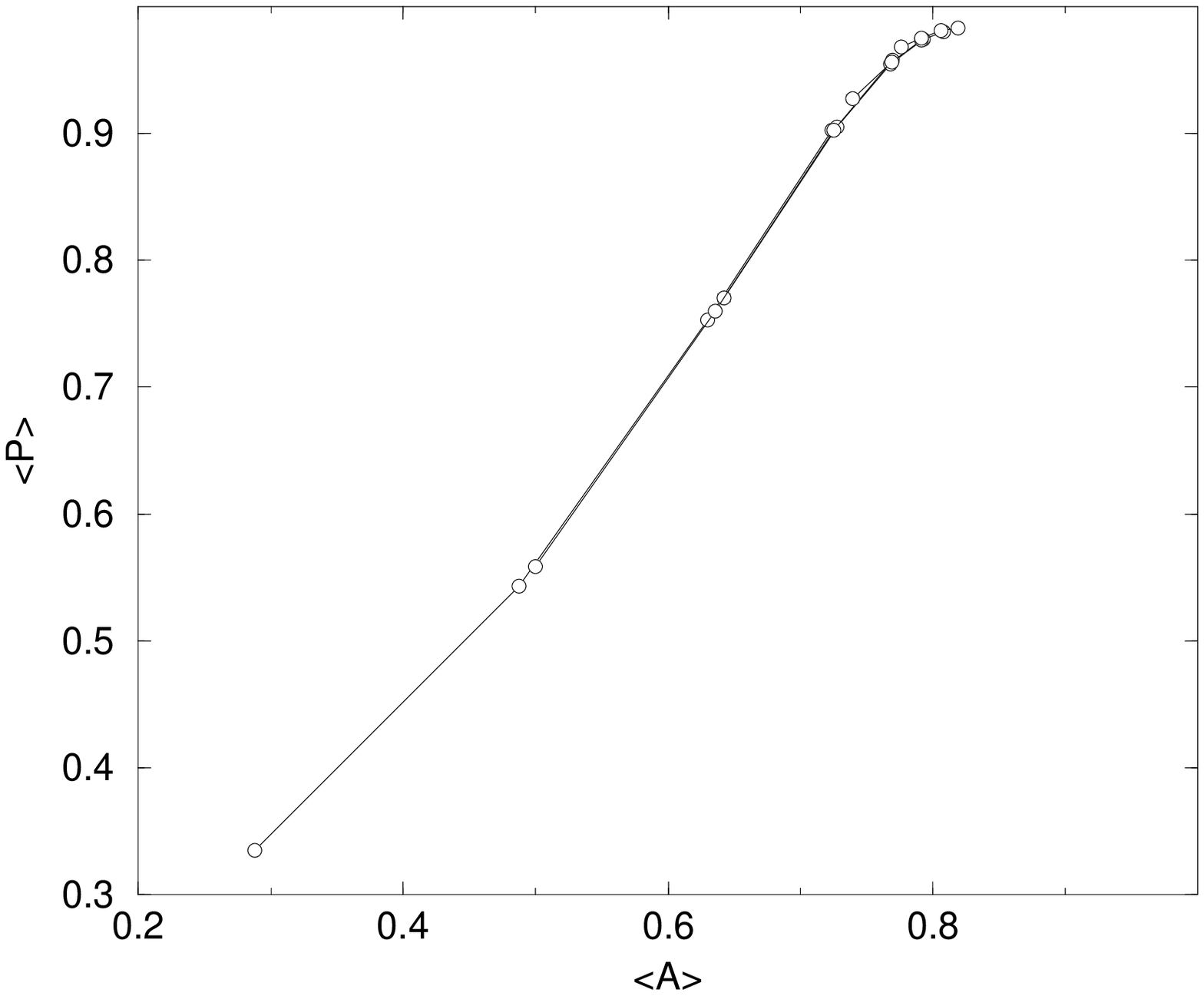,height=140mm}
\enc                         
\caption{\label{f3}\small The scaling flow built up from the RG flow segments 
of set1 given in table \ref{t1}. Each segment consists of four points corresponding to
the $(\la A,\ra,\la P,\ra)$ values on an $L=8$ lattice blocked for given couplings from lattices with
$L=64,128,256,512$, respectively. The segments are adjusted so that they overlay each other, 
and it can be seen from this figure, and from table \ref{t1}, that the points
on different segments coincide very well indicating that scaling holds. There are
two errant points corresponding to the two largest $\gb$ values and $L=512$.}
\enf

\bef      
\bec
\epsfig{file=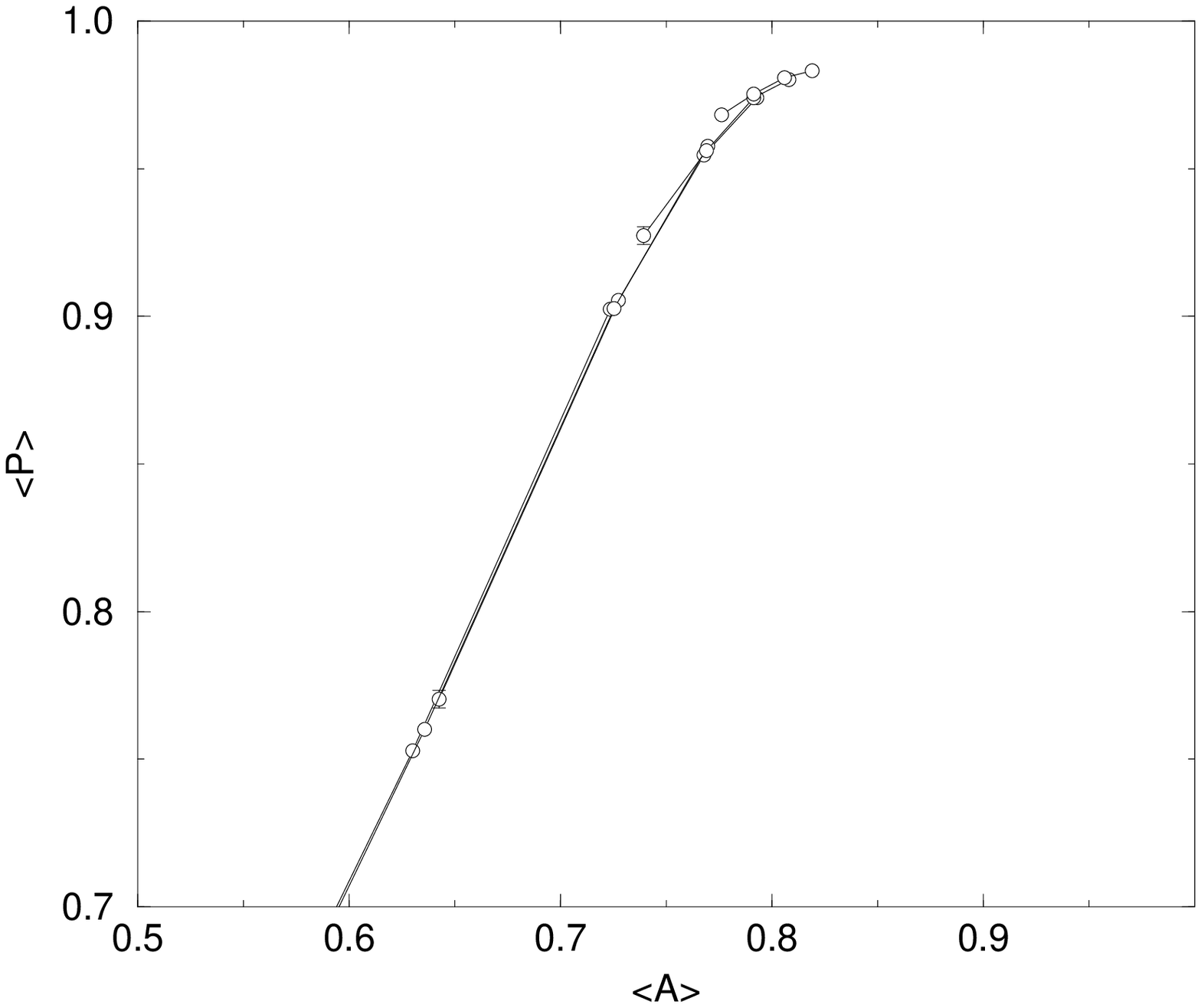,height=140mm}
\enc
\caption{\label{f4}\small Detail of figure \ref{f3}. Caption as for figure \ref{f3}}
\enf      

\bef
\bec
\epsfig{file=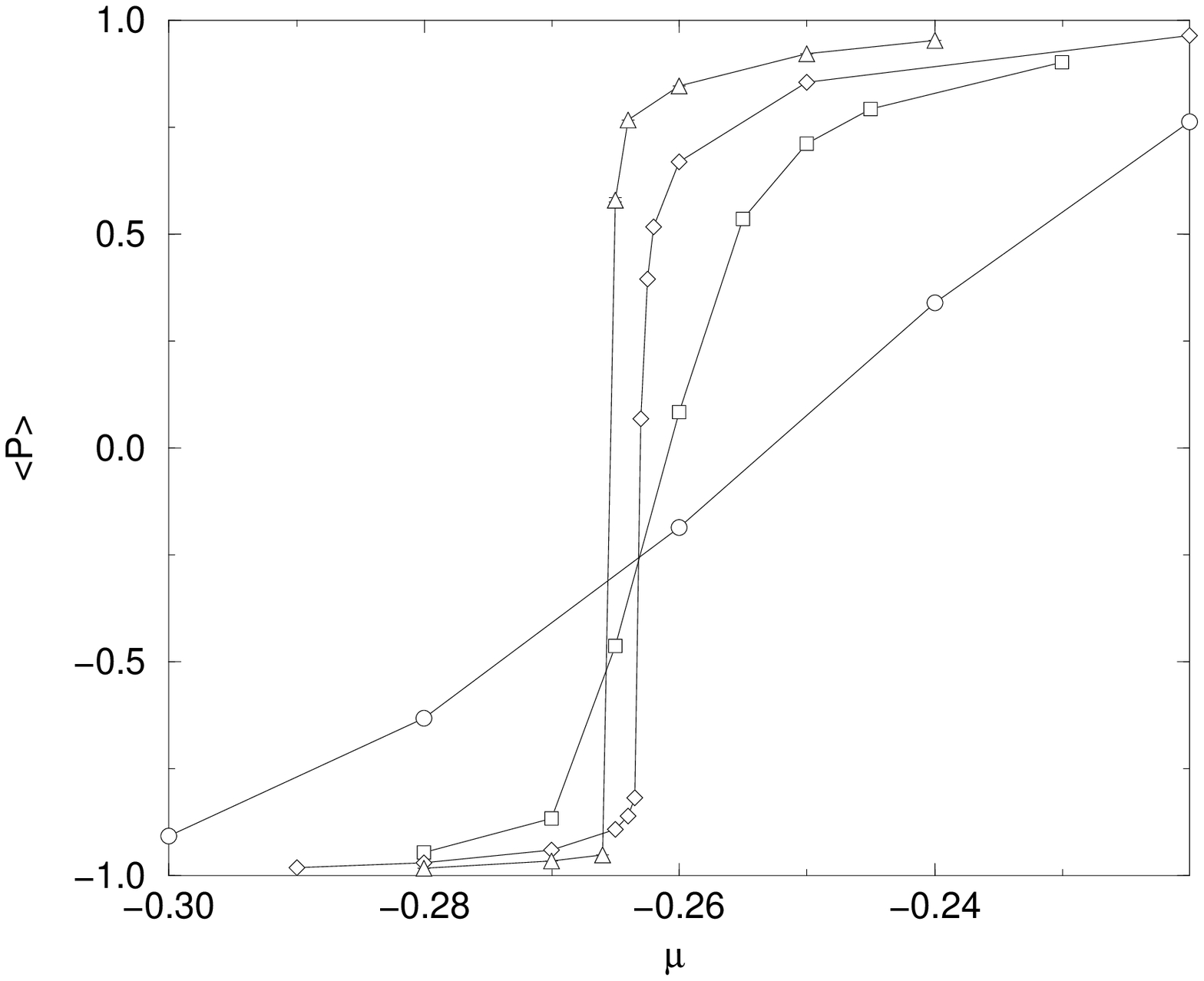,height=140mm}
\enc
\caption{\label{f7}\small $\PE$ defined by eq. \ref{AP} as a function of $\mu$
for $\gb=6.0~(\bigcirc), 7.0~(\Box)$, $7.5~(\Diamond), 8.0~(\bigtriangleup)$~.}
\enf

\bef
\bec
\epsfig{file=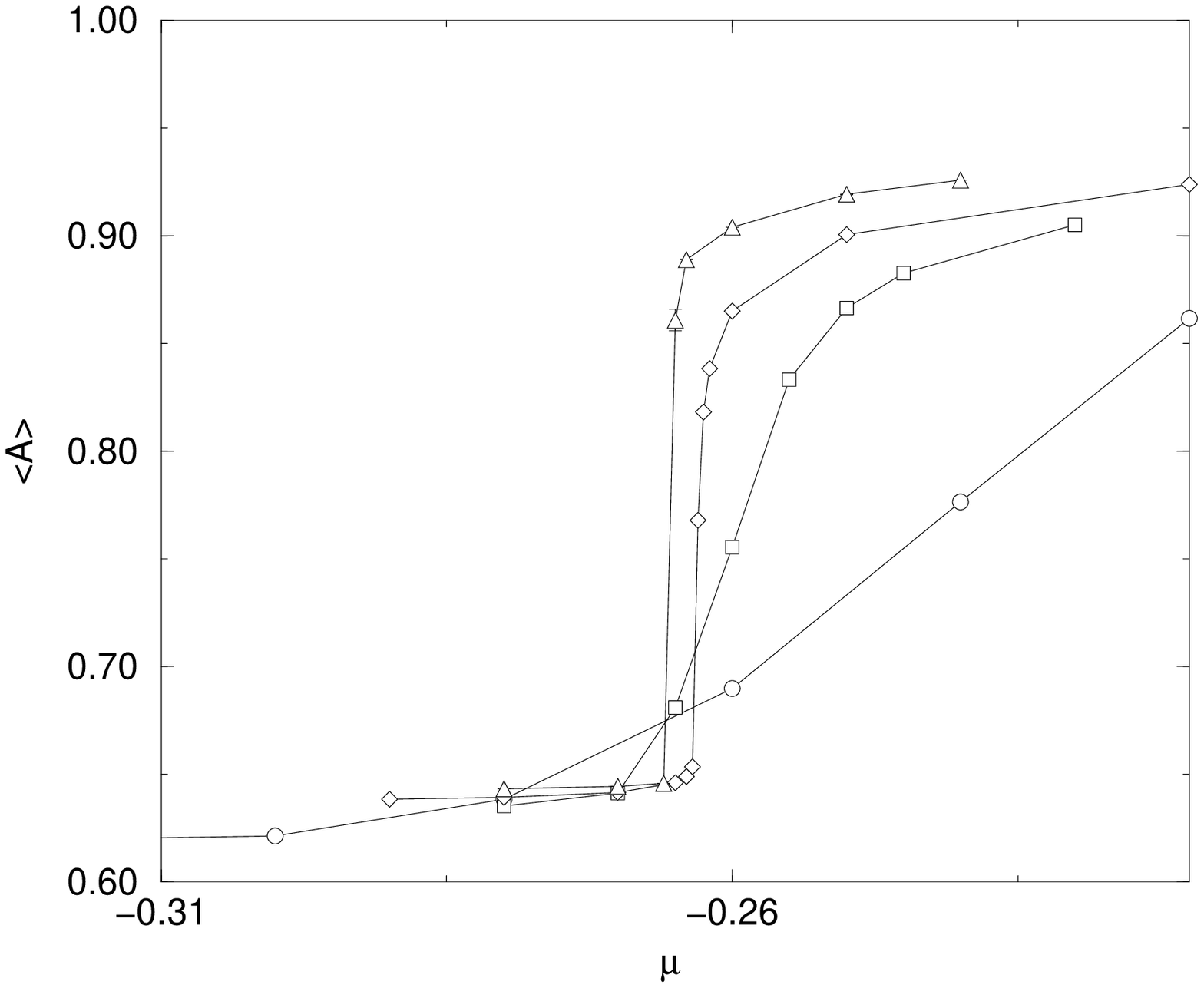,height=140mm}
\enc
\caption{\label{f8}\small $\AE$ defined by eq. \ref{AP} as a function of $\mu$
for $\gb=6.0~(\bigcirc), 7.0~(\Box)$, $7.5~(\Diamond), 8.0~(\bigtriangleup)$~.}
\enf

\bef
\bec
\epsfig{file=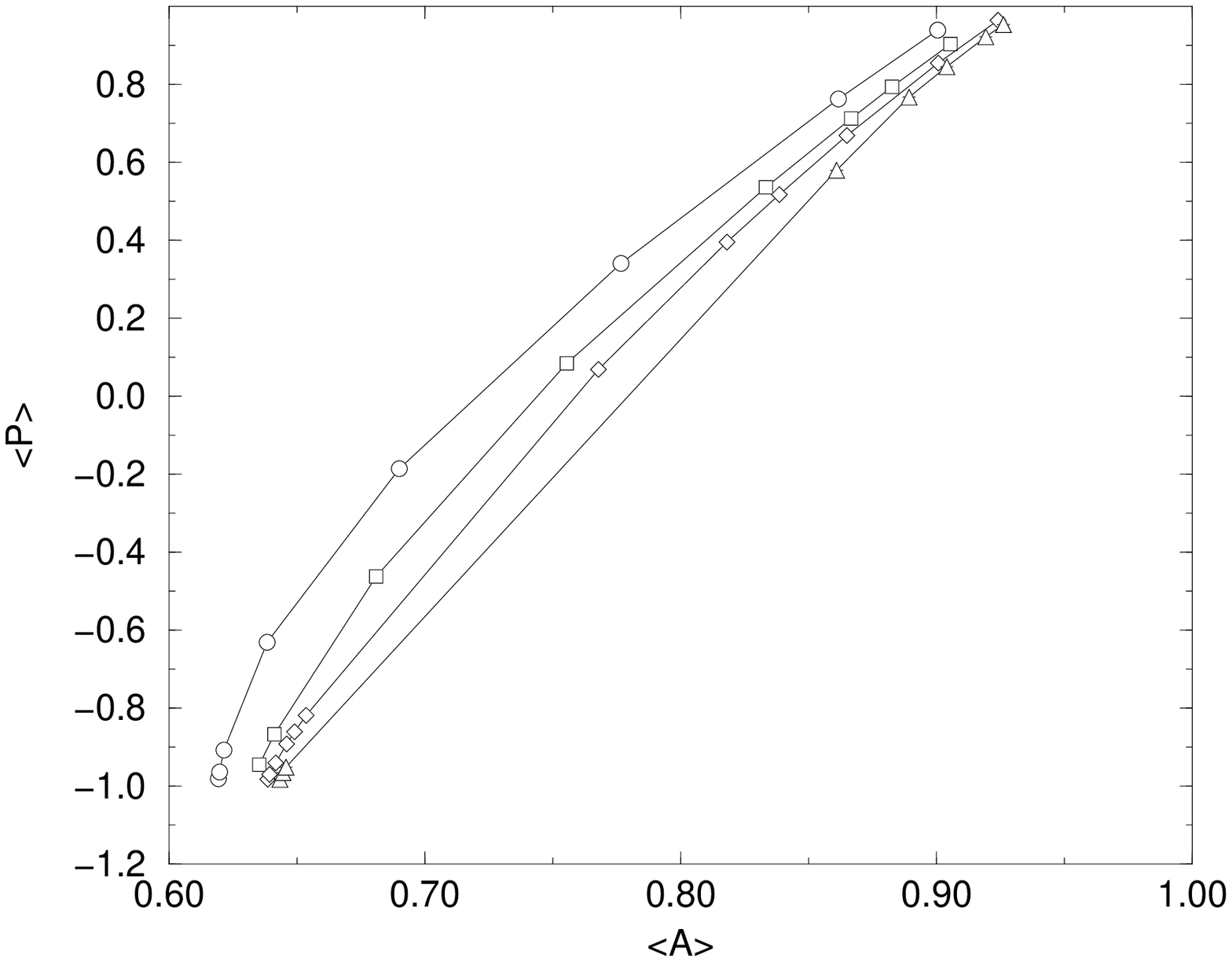,height=140mm}
\enc
\caption{\label{f9}\small $\AE$ versus $\PE$, defined by eq. \ref{AP}, for
$\gb=6.0~(\bigcirc), 7.0~(\Box), 7.5~(\Diamond), 8.0~(\bigtriangleup)$~.}
\enf

\bef                                                                                
\bec                                                                                              
\epsfig{file=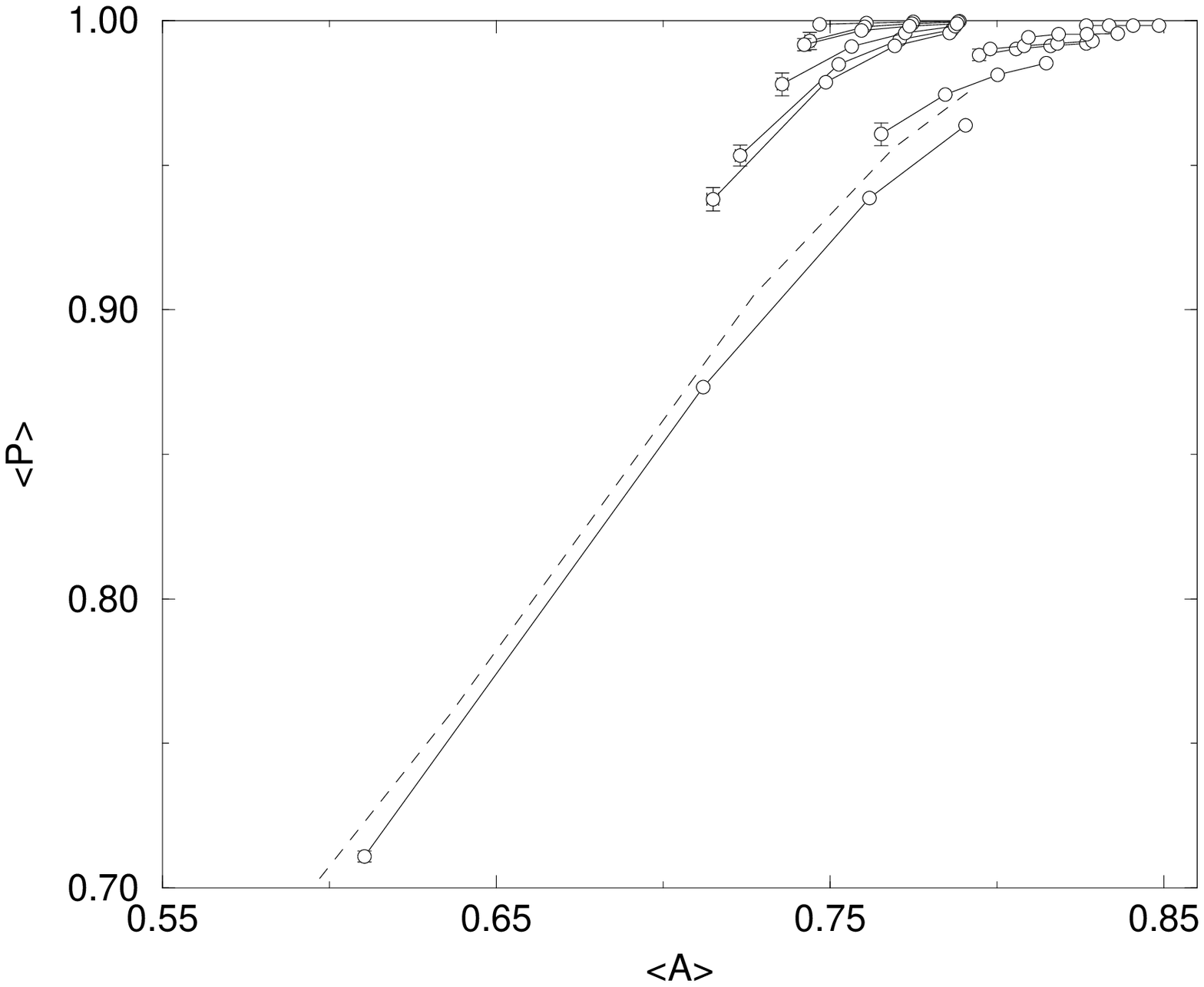,height=140mm}
\enc                                                                         
\caption{\label{f5}\small Two examples sets of flows, sets 3 and 4, showing cross-over 
of flows. Set 3 lies to the left of set 4 and the $(\gb,\,\mu)$ values corresponding to
these sets are given in table \ref{t3}. Set 4 is for various $\gb$ in the $RP^2$ model
($\mu=0$) in the range $\gb=3.9$ to $4.5$. There is rapid variation in the renormalized 
$(\la A,\ra,\la P,\ra)$ values for small changes in $\gb$ indicating a narrow cross-over from a region
of high to low renormalized vorticity. For $\gb \sim 4$ the $RP^2$ flows closely follow
the scaling flow of figure \ref{f3}, shown here as the dashed curve, implying 
scaling will apparently hold until, for larger $\gb$, the cross-over occurs.} 
\enf  

\bef                                                                                
\bec                                                                                              
\epsfig{file=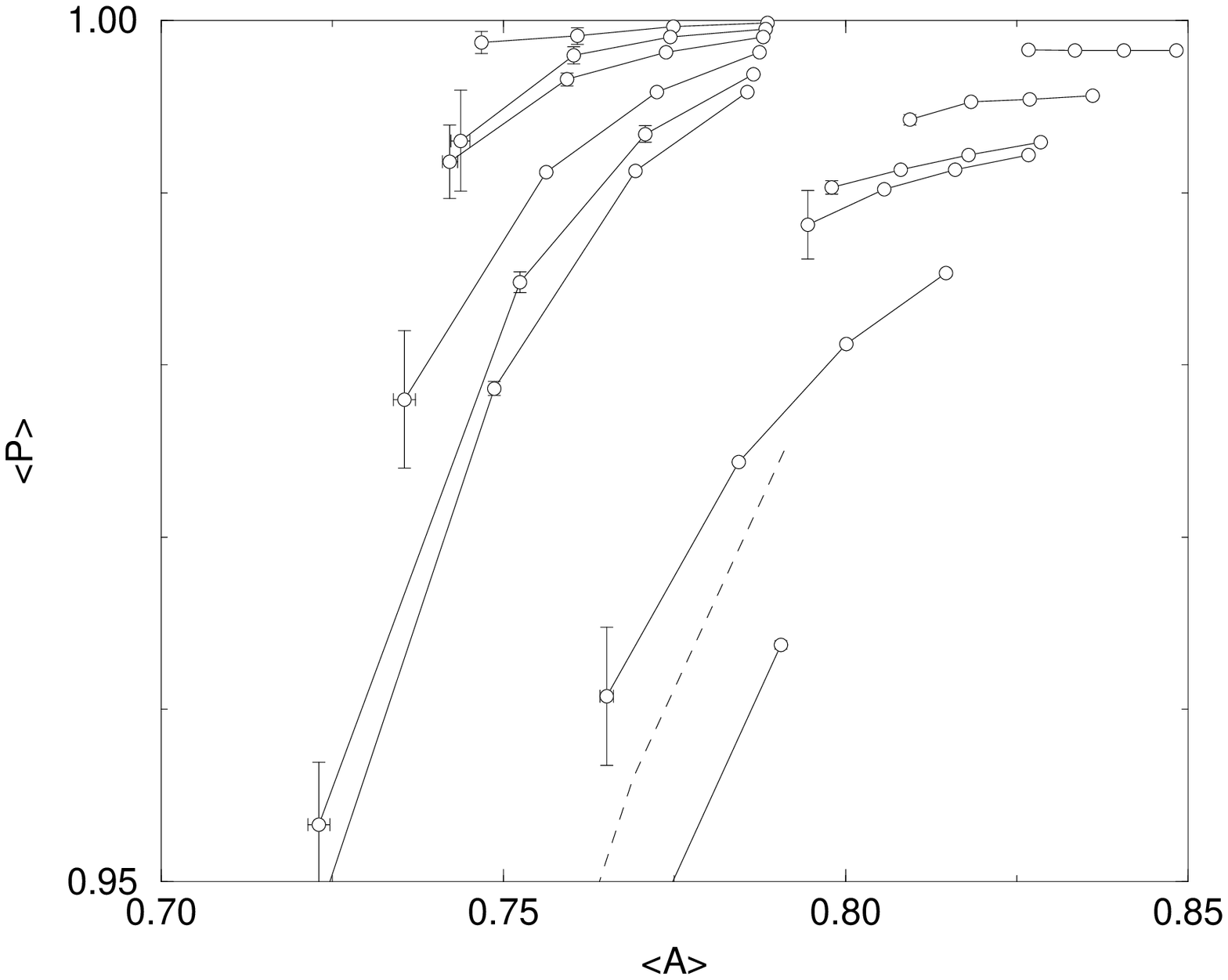,height=140mm}
\enc                                                                         
\caption{\label{f6}\small Detail of figure \ref{f5}. Caption as for figure \ref{f5}}
\enf  

\bef
\bec
\epsfig{file=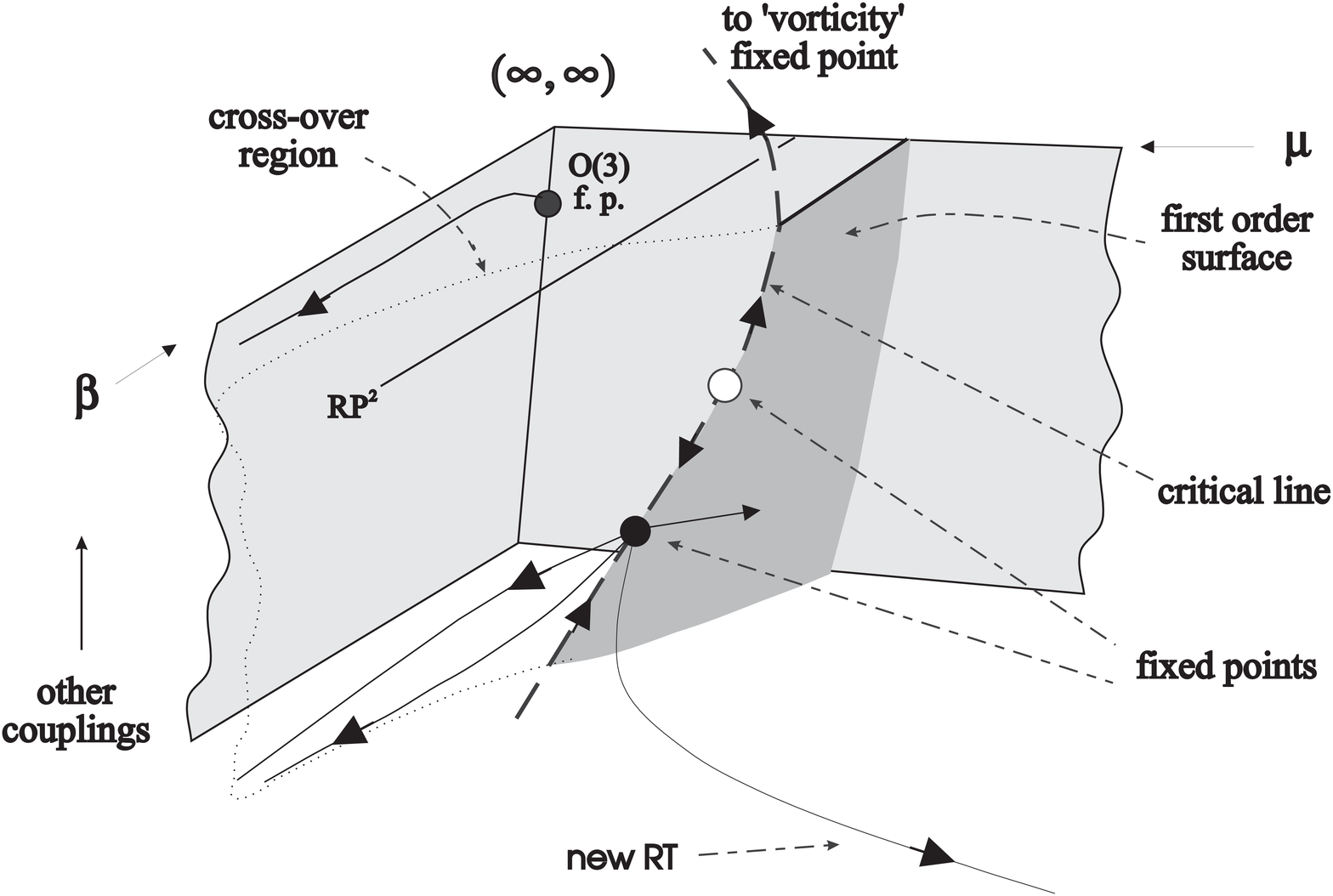,height=90mm}
\enc
\caption{\label{f10}\small An artist's impression of the RG flows consistent with the simulation results.
The $O(3)$ fixed point controls the continuum limit of $RP^2$ $(\mu=0)$ and neighbouring theories.
A line of critical points terminates the first-order surface and defines new continuum limits
characterized by  non-zero continuum vorticity.  The observed renormalized trajectory is shown 
associated with a new fixed point in the critical surface. There are indications that this 
fixed point does not control the second-order transition terminating the observed first-order line 
in the $(\gb,\mu)$ plane, and the scenario presented here is consistent with this view. The 
cross-over region is the neighbourhood of the surface shown with dotted outline. }
\enf

\end{document}

%% file: head.tex
\def\benum{\begin{enumerate}}
\def\eenum{\end{enumerate}}

\def\ben{\begin{equation}}
\def\ba{\begin{array}}
\def\bea{\begin{eqnarray}}
\def\bec{\begin{center}}

\def\een{\end{equation}}
\def\eea{\end{eqnarray}}
\def\enc{\end{center}}
\def\ea{\end{array}}
\def\btab{\begin{table}}
\def\btabu{\begin{tabular}}
\def\etab{\end{table}}
\def\etabu{\end{tabular}}
\def\bit{\begin{itemize}}
\def\eit{\end{itemize}}
\def\bef{\begin{figure}[htb]}
\def\befh{\begin{figure}[!h!]}
\def\enf{\end{figure}}

\def\Sc{\scriptstyle}

\def\la{\langle}
\def\ra{\rangle}
\def\d{\partial}
\def\a{\alpha}

\def\bphi{\mbox{\boldmath $\phi$}}
\def\bPhi{\mbox{\boldmath $\Phi$}}
\def\gb{\beta}

\def\g{\gamma}
\def\L{\Lambda}

\def\s{\sigma}

\def\V{{\cal V}}

\def\b1{{\bf 1}}

\def\bM{{\mbox{\boldmath $M$}}}

\def\bR{{\mbox{\boldmath $R$}}}
\def\bS{{\mbox{\boldmath $S$}}}

\def\bx{{\mbox{\boldmath $x$}}}
\def\by{{\mbox{\boldmath $y$}}}
\def\sbx{{\mbox{\boldmath $\Sc x$}}}
\def\sby{{\mbox{\boldmath $\Sc y$}}}

\def\bq{{\mbox{\boldmath $q$}}}

\def\sbmu{{\mbox{\boldmath $\Sc\mu$}}}
\def\sbnu{{\mbox{\boldmath $\Sc\nu$}}}
\def\bmu{{\mbox{\boldmath $\mu$}}}
\def\bnu{{\mbox{\boldmath $\nu$}}}
\def\k{{\kappa}}

\def\nn{\nonumber}
\def\bb{\left(}
\def\eb{\right)}
\def\bs{{\bf s}}

\def\AE{\la A \ra}
\def\PE{\la P \ra}
\newcommand{\name}{\arabic{section}}
\newcommand{\newsection}[1]{\section{#1}\renewcommand{\theequation}
                              {\name.\arabic{equation}}
                            \setcounter{equation}{0}}
\newcommand{\newsubsection}[1]{\subsection{#1}\renewcommand{\theequation}
                               {\name.\arabic{subsection}.\arabic{equation}}
                               \setcounter{equation}{0}}

\def\MS{{\overline{MS}}}

\def\bs{{\bf s}}

\def\rb{\raisebox{3mm}[0pt]}
\def\colsep{\setlength{\arraycolsep}{0pt}}
\usepackage{epsfig,latexsym}
\def\bem{\begin{minipage}}
\def\enm{\end{minipage}}

%% file: rpo3.bbl
\begin{thebibliography}{10}

\bibitem{SSD}
S.~Solomon et~al.
\newblock {\em Phys. Lett.}, 112B:373--378, 1982.

\bibitem{caea}
S.~Caracciolo et~al.
\newblock {\em Nucl. Phys.}, B Proc. Suppl. 30:815, 1993.

\bibitem{haho}
M.~Hasenbusch and R.R. Horgan.
\newblock {\em Phys. Rev.}, D53:5075--5089, 1996.

\bibitem{Sokal}
S.~Caracciolo et~al.
\newblock {\em Phys. Rev. Lett.}, 71:3906--3909, 1993.

\bibitem{holl}
T.~Hollowood.
\newblock {\em Phys. Lett.}, B329:450, 1994.

\bibitem{luea}
M.~L\"uscher, P.~Weisz, and U.~Wolff.
\newblock {\em Nucl. Phys.}, B359:221, 1991.

\bibitem{Nieder}
F.~Niedermayer et~al.
\newblock {\em Phys. Rev.}, D53:5918--5923, 1996.

\bibitem{Has}
M~Hasenbusch.
\newblock {\em Phys. Rev}, D53:3445--3450, 1996.

\bibitem{NN}
B.~Nienhuis and N.~Nauenberg.
\newblock {\em Phys. Rev. Lett.}, 35:477--479, 1975.

\bibitem{haha}
A.~Hasenfratz and P.~Hasenfratz.
\newblock {\em Nucl. Phys.}, B295:1, 1988.

\bibitem{goea}
A.P. Gottlob, M.~Hasenbusch, and K.~Pinn.
\newblock {\em Phys. Rev.}, D54:1736--1747, 1996.

\bibitem{hoho}
A.~Hoch and R.R. Horgan.
\newblock {\em Nucl. Phys.}, B380:337--366, 1992.

\bibitem{siea}
Seiler E. et~al.
\newblock {\em Nucl. Phys.}, B305:623, 1988.

\bibitem{jana}
Janke W. and Nather K.
\newblock {\em Phys. Lett}, A157:11, 1991.

\bibitem{kuzu}
H.~Kunz and G.~Zumbach.
\newblock {\em Phys. Rev.}, B46:662, 1992.

\bibitem{hupo}
K.~Huang and J~Polonyi.
\newblock {\em Int. J. Mod. Phys.}, A6:409--430, 1991.

\end{thebibliography}
